\documentclass[Afour,sageh,times]{sagej}

\usepackage{moreverb,url}
\usepackage[colorlinks,bookmarksopen,bookmarksnumbered,citecolor=blue,urlcolor=blue]{hyperref}
\newcommand\BibTeX{{\rmfamily B\kern-.05em \textsc{i\kern-.025em b}\kern-.08em
T\kern-.1667em\lower.7ex\hbox{E}\kern-.125emX}}
\def\volumeyear{2025}
\begin{document}
\title{Explosive Growth in Large-Scale Collaboration Networks}
\runninghead{Williams and Chen}
\author{Peter R. Williams\affilnum{1,2} and Zhan Chen\affilnum{3}}
\affiliation{\affilnum{1}Rinna KK, Tokyo, Japan\\\affilnum{2}Independent Researcher\\\affilnum{3}Microsoft Japan, Tokyo, Japan}
\email{prw20042004@yahoo.co.uk}

\begin{abstract}
We analyse the evolution of two large collaboration networks: the Microsoft Academic Graph (1800-2020) and Internet Movie Database (1900-2020), comprising $2.72 \times 10^8$ and $1.88 \times 10^6$ nodes respectively. The networks show super-linear growth, with node counts following power laws $N(t) \propto t^{\alpha}$ where $\alpha = 2.3$ increasing to $3.1$ after 1950 (MAG) and $\alpha = 1.8$ (IMDb). Node and edge processes maintain stable but noisy timescale ratios ($\tau_N/\tau_E \approx 2.8 \pm 0.3$ MAG, $2.3 \pm 0.2$ IMDb). The probability of waiting a time $t$ between successive collaborations was found to be scale-free, $P(t) \propto t^{-\gamma}$, with indices evolving from $\gamma \approx 2.3$ to $1.6$ (MAG) and $2.6$ to $2.1$ (IMDb). Academic collaboration sizes increased from $1.2$ to $5.8$ authors per paper, while entertainment collaborations remained more stable ($3.2$ to $4.5$ actors). These observations indicate that current network models might be enhanced by considering accelerating growth, coupled timescales, and environmental influence, while explaining stable local properties.
\end{abstract}

\maketitle

\section{Introduction}

Collaboration networks are a fundamental part of human society. These networks connect people through various types of relationships, from friendship and professional collaboration to online interactions. Network science provides mathematical tools to analyse these complex systems of connections \citep{newman2018}.

The structure and evolution of these networks have been studied extensively. Most understanding comes from examining networks at specific points in time. This snapshot approach has revealed several universal properties across different types of networks. Among the commonly observed properties are the small-world phenomenon \citep{watts1998}, the presence of power-law degree distributions \citep{barabasi2016}, and high clustering coefficients \citep{holme2019temporal}.

The small-world phenomenon describes two key network features that can be precisely quantified. The first is a short average path length $L$, defined as the mean number of steps along the shortest paths between all pairs of nodes. In most real networks, $L$ scales logarithmically with network size \citep{watts1998}. The second feature is a high clustering coefficient $C$, which measures the probability that two neighbours of a node are themselves connected. Typically, $C$ is orders of magnitude larger than would be expected in random networks of the same size \citep{newman2003}.

The degree distribution $P(k)$ of a network shows the probability that a randomly chosen node has $k$ connections. In many real networks, this distribution follows a power law: $P(k) \propto k^{-\gamma}$, where $\gamma$ typically falls between 2 and 3 \citep{broido2019}. Recent work has shown that identifying true power laws requires careful statistical analysis \citep{voitalov2019}: many networks previously thought to follow power laws may be better described by other heavy-tailed distributions \citep{seshadri2021}.

Recent theoretical work has highlighted the possibility of non-linear and explosive growth in networks. The concept of explosive percolation, introduced by \citet{achlioptas2009}, showed that networks can undergo sudden transitions in connectivity. This was extended by studies of preferential attachment with fitness \citep{bianconi2001,dorogovtsev2000}, which showed how node heterogeneity can lead to accelerated growth patterns. \citet{krioukov2010} showed how underlying geometric structures can influence network growth dynamics, potentially leading to different scaling properties.

Empirical studies have found evidence for non-linear growth in various contexts. \citet{leskovec2007} observed densification and faster-than-linear edge growth in several real networks. \citet{kumar2010} documented similar phenomena in online social networks. However, most empirical studies cover relatively short time periods, making it difficult to distinguish between transient acceleration and fundamental growth patterns.

Current models attempt to explain how these network properties emerge. The most widely accepted is preferential attachment \citep{barabasi1999}, where new nodes prefer to connect to highly connected existing nodes. This model has been refined extensively \citep{holme2019network}. Recent variations include fitness-based attachment \citep{kong2019}, temporal preferences \citep{ubaldi2020}, and memory effects \citep{zhang2021}.

Recent work has highlighted the importance of carefully examining assumptions about network growth patterns. \citet{leskovec2007} showed that many real networks show systematic deviation from constant growth rates, while \citet{mitzenmacher2004} provided theoretical foundations for understanding accelerating growth processes. The temporal aspects of network evolution have received increased attention, with \citet{karsai2011} and \citet{kossinets2006} highlighting the complex interplay between different timescales in social networks.

These models make several important but rarely tested assumptions:
\begin{enumerate}
    \item \textbf{Steady Growth}: Networks are assumed to grow at a constant or slowly varying rate. Recent work suggests this may not hold for online social networks \citep{ribeiro2014}, but long-term studies are rare.

    \item \textbf{Timescale Separation}: Node addition is assumed to happen much more slowly than edge addition. This assumption appears in most theoretical frameworks \citep{holme2019temporal} but lacks empirical validation.

    \item \textbf{System Isolation}: Networks are treated as isolated systems, ignoring external influences. Some recent studies suggest this may be an oversimplification \citep{karimi2018}.

    \item \textbf{Equilibrium Dynamics}: Many models assume the network reaches a quasi-steady state where structural properties stabilise \citep{krioukov2016}. This may not reflect reality in rapidly growing networks.
\end{enumerate}

Empirical studies of network evolution over time have been limited in both duration and scale. Most cover periods shorter than a decade \citep{mislove2008,kumar2010}. Even recent studies of large online social networks typically analyse networks with millions of nodes over a few years \citep{leskovec2020}. Studies of network evolution over decades are rare, with notable exceptions being small citation networks \citep{sinatra2015} and limited collaboration networks \citep{vazquez2018}.

Mathematical frameworks for analysing growing networks have advanced significantly. Recent developments include non-Markovian growth processes \citep{williams2019}, temporal network motifs \citep{paranjape2017}, and higher-order network structures \citep{benson2018}. However, these frameworks typically assume the growth assumptions listed above.

In this paper, we analyse two large collaboration networks over extended time periods: the Microsoft Academic Graph (MAG) covering 1800-2020, and the Internet Movie Database (IMDb) covering 1900-2020. These datasets provide an extensive historical perspective for studying network evolution:
\begin{enumerate}
    \item Temporal span: $>100$ years of continuous evolution,
    \item Scale: Hundreds of millions of nodes and edges,
    \item Quality: Rigorous data curation and validation,
    \item Completeness: Comprehensive coverage within their domains.
\end{enumerate}
Our analysis indicates patterns that differ from some current assumptions about network growth and appear consistent with recent theoretical work on accelerating networks. We found
\begin{enumerate}
\item Explosive rather than steady growth, with rates following power laws with exponents $>1$, similar to but more extreme than the densification observed by \citet{leskovec2007};
\item Similar timescales for node and edge processes (ratio $2.8 \pm 0.3$), contradicting the timescale separation assumption implicit in many models since \citet{barabasi1999};
\item Clear signatures of external influences on growth patterns, supporting the importance of temporal network theory as outlined by \citet{holme2015}; and
\item No evidence of approach to equilibrium even after centuries of evolution.
\end{enumerate}
These findings suggest that current models of network growth may need significant revision to represent real network evolution accurately. They also highlight the importance of empirical validation for theoretical assumptions, especially in complex systems like social networks.

The richness and complexity of these two networks, combined with the multiple novel findings, necessitates presenting our analysis across a series of focused papers. This first paper establishes the fundamental dynamics of network growth, examining the rates and patterns of node and edge evolution. The second paper investigates node and edge lifetime processes, identifying patterns in career longevity and analysing the temporal evolution of collaboration dynamics across different domains. In the third paper, we analyse network degree distributions, comparing Weibull and power-law models, and examine the potential role of structural constraints in network formation. The fourth paper studies the evolution of network complexity through entropy measures and small-world properties, analysing changes in hierarchical organisation and information flow over time. The final paper examines network responses to major historical events, analysing patterns of disruption and recovery in academic versus entertainment collaborations, and investigating the persistence of these effects. This division allows thorough treatment of each aspect while maintaining clear thematic focus. The present paper provides the foundation for this series by establishing the basic growth dynamics that underlie all other network properties.

The paper is organised as follows. The next section details the construction of our datasets from the Microsoft Academic Graph and Internet Movie Database, including careful consideration of temporal metadata and the development of robust measures for network growth and characteristic timescales. The following section presents quantitative analysis of network evolution, revealing unexpected patterns of super-linear growth, coupled timescales between node and edge processes, and distinct waiting time distributions that characterise professional collaboration rhythms. We then examine the theoretical implications of these findings in a discussion section, considering how they complement existing network models and suggest new directions for theory development that can incorporate sustained acceleration and environmental coupling while accounting for stable local properties. Finally, we summarise our key findings and outlines critical directions for future research, emphasising the need to extend this analysis to other forms of social networks and develop more sophisticated theoretical frameworks.

This paper focuses on establishing empirical patterns in the long-term evolution of collaboration networks, providing a foundation for subsequent analyses. Our primary objectives are to: (1) quantify the fundamental growth dynamics of nodes and edges in both networks over century-long timescales, (2) measure and analyze the characteristic timescales of network processes, and (3) document patterns in collaboration waiting times and team sizes. We adopt an observational approach, remain model-agnostic, presenting detailed measurements of network properties and their evolution over time. While we discuss our findings in the context of existing theoretical frameworks, our goal is to establish robust empirical patterns rather than to advocate for particular models or mechanisms. This approach allows the data to speak for itself, providing a foundation for future theoretical work that can build on these observations. Other important aspects of these networks, including degree distributions, information flow, and responses to historical events, will be examined in companion papers. By maintaining this empirical focus, we aim to establish clear baseline observations about how large-scale collaboration networks evolve over extended periods.

\section{Data and Methods}

\subsection{Methodological Approach}

We employ an empirical, observation-based approach to network analysis, measuring network properties and their evolution directly, rather than through theoretical models. While we may fit mathematical functions to our data in future papers, these serve as descriptive tools rather than theoretical explanations. This approach aims to identify empirical patterns in network evolution that can inform future theoretical work. Our focus on observation helps complement existing network theories, which often drew from shorter-term studies. Throughout our analysis, we attempt to separate observations from theoretical interpretations.

\subsection{Network Construction and Temporal Analysis}

This study analyses two large collaboration networks: the Microsoft Academic Graph (MAG) \citep{magweb, sinha2015} and the Internet Movie Database (IMDb) \citep{imdbweb}. Both datasets are freely available and provide comprehensive coverage within their respective domains, as well as careful consideration of name disambiguation and data quality. 

Each network was constructed as an undirected graph evolving in time. For both networks, nodes represent individual contributors (authors in MAG, actors in IMDb), with edges representing projects (papers or movies) on which the nodes were collaborating. Each edge in the network was labelled with both its creation and removal time, in order to track the temporal evolution of projects and node collaborations.

The MAG dataset comprises $\sim$80 Gb of text data covering scientific publications from 1800 to 2020. After processing, our analysis includes:
\begin{itemize}
    \item $2.72 \times 10^8$ nodes (authors)
    \item $2.64 \times 10^8$ papers
    \item $\sim 1.8 \times 10^9$ edges
\end{itemize}
A small fraction of publications ($65,111$) lacking publication dates were excluded from the analysis.

The IMDb dataset comprises $\sim$10 Gb of text data covering films from 1900 to 2020. After processing, the co-star network includes:
\begin{itemize}
    \item $1.88 \times 10^6$ nodes (actors and actresses)
    \item $6.34 \times 10^5$ movies
    \item $\sim 1.8 \times 10^6$ edges
\end{itemize}

Network edges were constructed from documented collaborations in papers and movies. For each publication or film with $n$ contributors, edges connecting every possible pair of contributors were generated, producing $n(n-1)/2$ new edges. This approach yielded a fully connected subgraph for each collaboration. Precise publication and release dates, obtained from the source data, determined edge removal times. Each edge in the network represents a distinct collaborative project and carries unique temporal metadata, including both start and end timestamps. Multiple edges may exist between any node pair $(i,j)$, reflecting repeated collaborations between contributors across different projects.

The analysis of network evolution called for a temporal framework that carefully addressed collaboration duration. While project completion times $t_{\text{end}}$ were explicitly documented through publication or release dates, project initiation times $t_{\text{start}}$ required estimation. The primary analysis employed a simplified model with a fixed collaboration duration of $\tau_{\text{project}} = 2$ years. Under this model, edges were added to the network at time $t_{\text{start}} = t_{\text{end}} - \tau_{\text{project}}$ and removed at $t_{\text{end}}$. Two nodes $i$ and $j$ were defined as collaborating at time $t$ if at least one edge $e_{ij}(t)$ existed between them at that moment.

The stability of this temporal framework was examined through sensitivity analyses. Alternative duration models were tested, including fixed periods ranging from $\tau_{\text{min}} = 3$ months to $\tau_{\text{max}} = 4$ years and Gaussian-distributed random durations with mean $\mu_{\tau}$ and standard deviation $\sigma_{\tau}$. The effects of varying temporal resolution were examined through different time-binning windows $\Delta t \in [3\text{ months}, 4\text{ years}]$, with annual bins ($\Delta t = 1\text{ year}$) selected for the final analysis. The key findings remained consistent across these variations, as they primarily concerned trends operating on timescales $T \gg \tau_{\text{project}}$.

\subsection{Growth Measures}

Five fundamental quantities were measured annually as a function of time $t$,
\paragraph{Node Dynamics:}
\begin{itemize}
    \item $N_{\text{new}}(t)$: new nodes added,
    \item $N_{\text{total}}(t)$: total number of nodes possessing edges at time $t$,
    \item $f_{\text{new}}(t) = N_{\text{new}}(t)/N_{\text{total}}(t)$: fraction of new nodes.
\end{itemize}
\paragraph{Edge Dynamics:}
\begin{itemize}
    \item $E_{\text{new}}(t)$: new edges added,
    \item $E_{\text{total}}(t)$: total active edges.
\end{itemize}
For each process, we computed a characteristic timescale
\begin{equation}
    \tau_{\text{process}} = \frac{\text{Total quantity}}{\text{Rate of change}}.
\end{equation}
These timescales enabled direct comparison between node and edge dynamics.

\section{Results}

\begin{figure*}[htb]
  \centering
  \begin{tabular}{cc}
    \includegraphics[width=0.48\textwidth]{"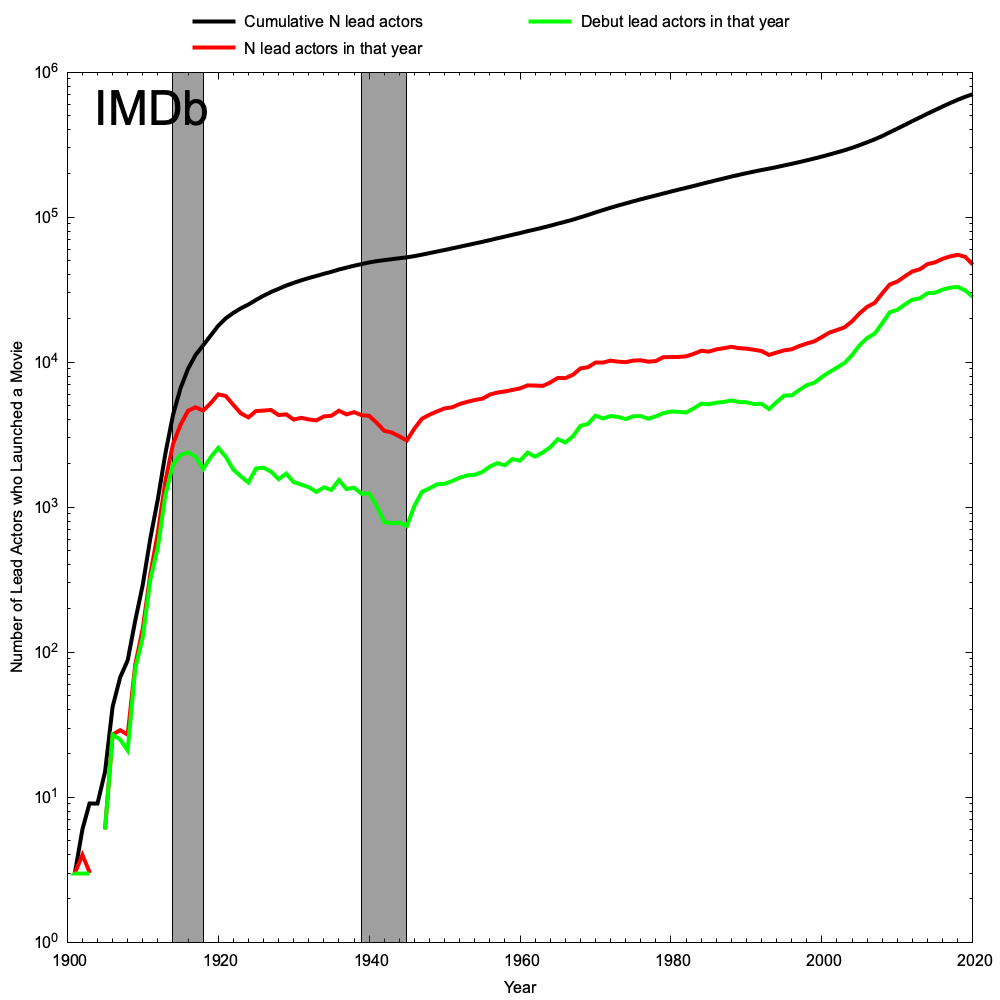"} &
    \includegraphics[width=0.48\textwidth]{"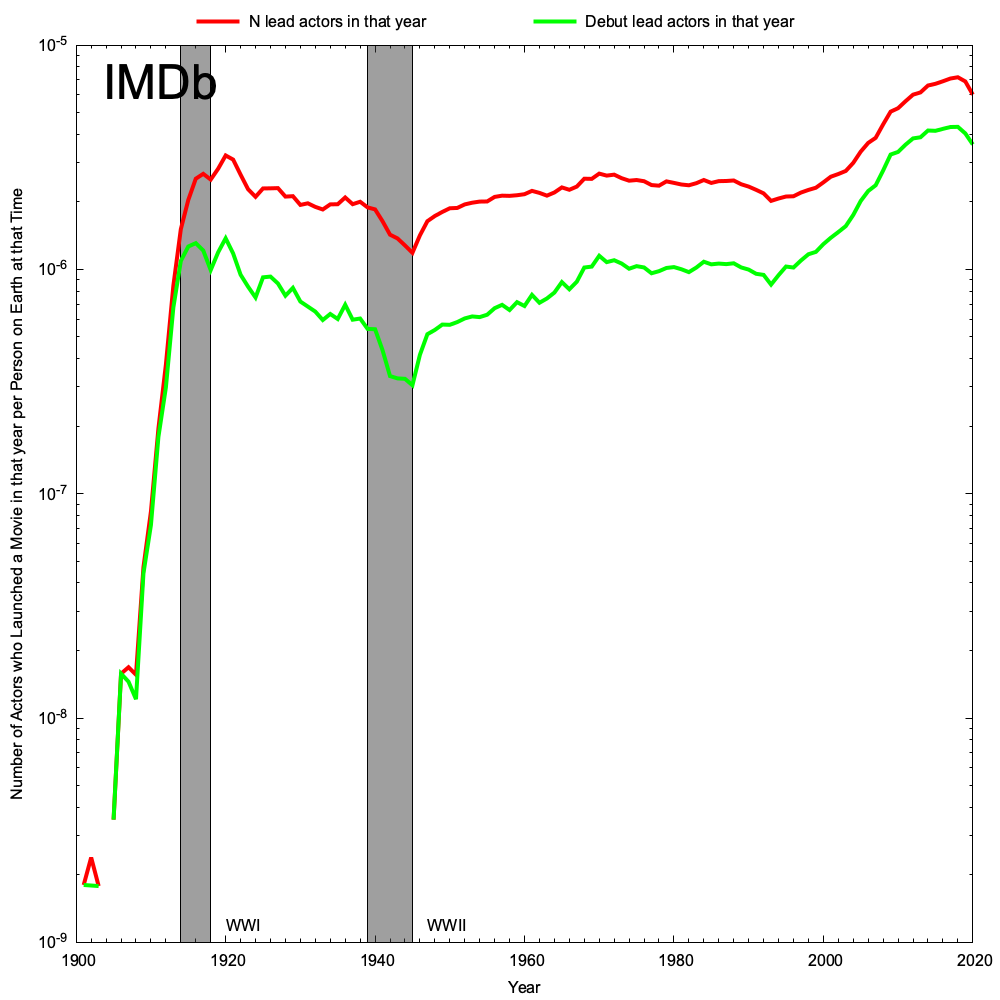"} \\
    \includegraphics[width=0.48\textwidth]{"nauthors.png"} &
    \includegraphics[width=0.48\textwidth]{"nauthors_population_adjusted.png"}
  \end{tabular}
  \caption{Evolution of node counts in the MAG (top) and IMDb (bottom) networks on logarithmic scales. Left panels show absolute counts: cumulative total nodes (black line), nodes active with new edges that year (red line), and new nodes joining that year (green line). Right panels show the same data normalised by the contemporary world population. Grey bands indicate major historical events: La Belle Epoque (1890-1914), World War I (1914-1918), and World War II (1939-1945). Population data post-1950 uses official UN records; earlier values are linearly interpolated between historical estimates. Note the distinct change in MAG growth rate around 1950 and the different sensitivities to historical events between networks.}
  \label{node_count}
\end{figure*}

\subsection{Explosive Network Growth}

Both the MAG and IMDb networks show substantial growth with increasing rates of expansion. Figure \ref{node_count} shows three key measures of network evolution: the cumulative total number of nodes (black line), the number of nodes with new edges in each year (red line), and the number of new nodes joining the network (green line). For both networks, these measures show dramatic deviations from steady growth models, with several distinct characteristics.

To parameterize the observed super-linear growth, we fit power-law functions to the cumulative node counts. It is important to note that these power-law fits serve primarily as descriptive approximations to capture the overall scaling trend, rather than as definitive claims of strict power-law behavior across all timescales.

The MAG network appears to show two different growth patterns, with a transition occurring around 1950. In the pre-1950 era, the cumulative node count follows a power-law $N(t) \propto t^{\alpha_1}$ with $\alpha_1 = 2.3 \pm 0.1$. After 1950, the growth accelerates, following $N(t) \propto t^{\alpha_2}$ with $\alpha_2 = 3.1 \pm 0.1$. This acceleration may indicate a change in network dynamics, coinciding with the post-war expansion of scientific research. The annual count of new nodes (green line) shows similar acceleration, but with pronounced sensitivity to historical events, marked by grey bands in Figure \ref{node_count}. These events include World War I (1914-1918), and World War II (1939-1945), each producing significant deviations from the background growth trend. These external pressures on the network and their effects are explored fully in a future paper in this series.

The IMDb network shows qualitatively similar but quantitatively more moderate acceleration. From approximately 300 new actors per year in 1950, the network expands to accommodate roughly 3000 new actors annually by 2020. The overall growth pattern for cumulative nodes follows $N(t) \propto t^{\beta}$ with $\beta = 1.8 \pm 0.1$. The annual new node count (green line) shows less pronounced sensitivity to historical events compared to the MAG network, though the impact of World War II is still visible.

\begin{figure*}[htb]
  \centering
  \begin{tabular}{cc}
    \includegraphics[width=0.48\textwidth]{"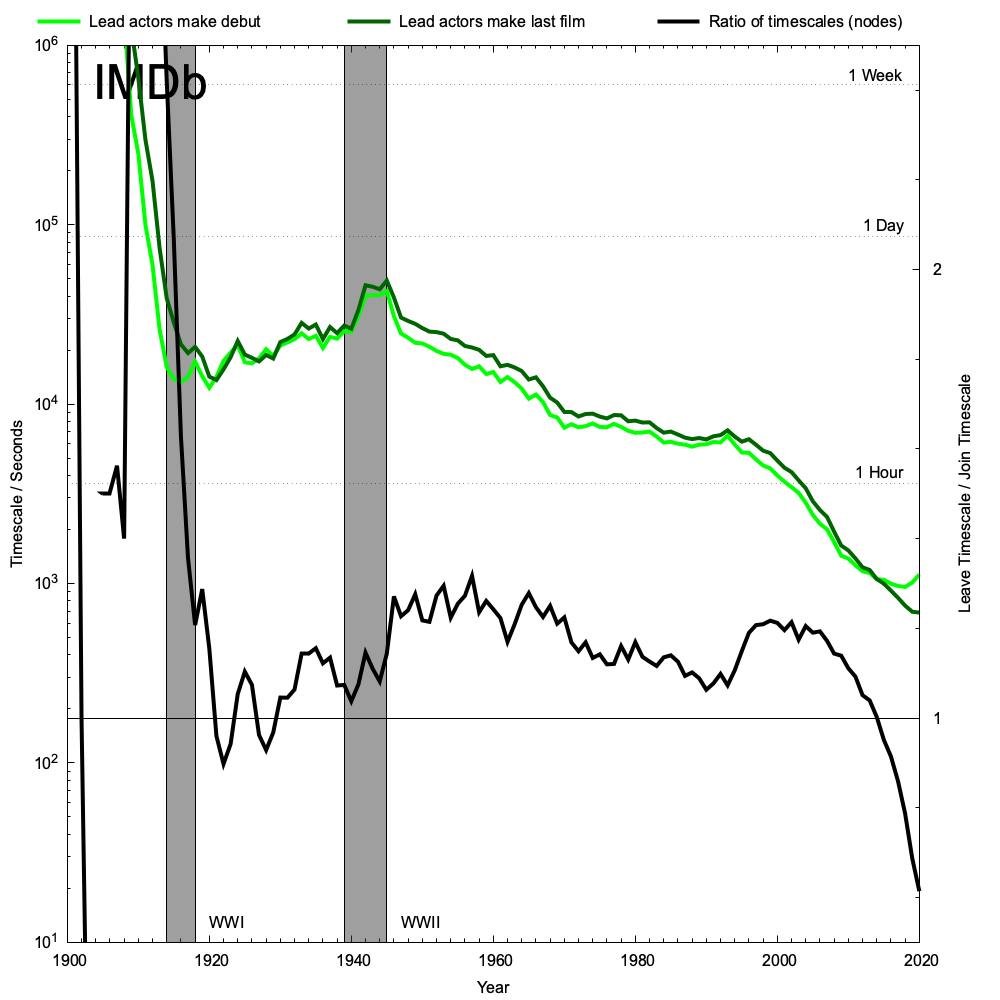"} &
    \includegraphics[width=0.48\textwidth]{"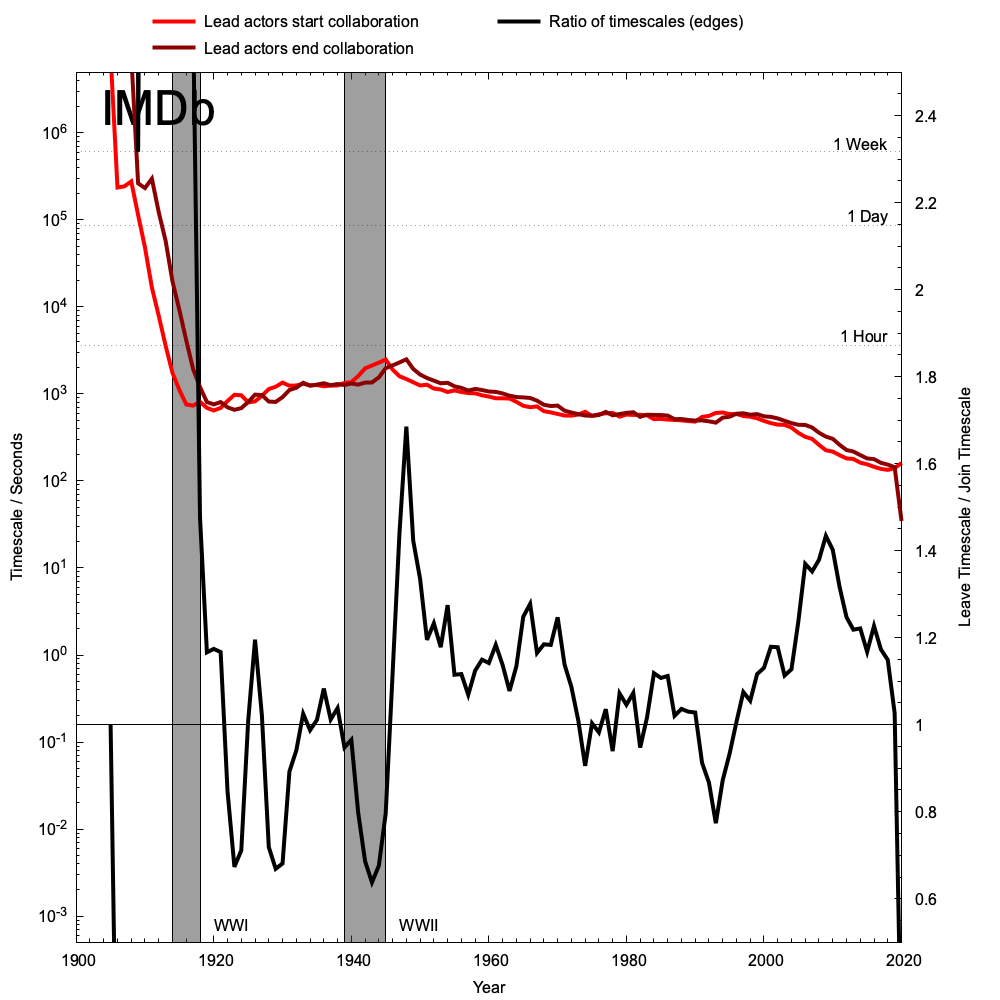"} \\
    \includegraphics[width=0.48\textwidth]{"timescales-nodes.png"} &
    \includegraphics[width=0.48\textwidth]{"timescales-edges.png"}
  \end{tabular}
  \caption{Characteristic timescales of network processes, shown on logarithmic scales. Top panels: MAG network timescales; Bottom panels: IMDb network timescales. Left panels show node timescales: addition (black) and removal (red). Right panels show edge timescales: addition (black) and removal (red). Timescales were computed as the ratio of total quantity to its rate of change. Note the parallel evolution of timescales within each network despite their different absolute values, and the stability of their ratios over centuries of evolution.}
  \label{timescales}
\end{figure*}

Population-adjusted measures (right panels) reveal that this explosive growth far exceeds demographic expansion. While world population increased by a factor of approximately 8 between 1800 and 2020, the MAG network grew by a factor of nearly 10,000 during the same period. Even after population adjustment, both networks maintain super-linear growth rates. The adjusted data shows that the per capita rate of joining these networks has increased by more than three orders of magnitude, indicating fundamental changes in the underlying social and professional structures driving collaboration.

\subsection{Analysis of Process Timescales}

Our analysis suggests quantitative relationships between node and edge process timescales that differ from traditional assumptions. Figure \ref{timescales} presents four key timescale measurements: node addition and removal timescales (left panels), and edge addition and removal timescales (right panels), for both networks. These characteristic timescales were defined for nodes and edges as
\begin{equation}
  \tau_N = \frac{N_{\text{total}}}{dN/dt}
\end{equation}
and
\begin{equation}
  \tau_E = \frac{E_{\text{total}}}{dE/dt},
\end{equation}
where the time derivatives were estimated with difference equations. These ratios measure the time required for significant network changes relative to total size.

The MAG network timescales (top panels) show systematic evolution over two centuries. The node addition timescale (black line, top left) decreases from $\sim$20 years in 1800 to $\sim$3 years by 2000, following a power law $\tau_N \propto t^{-0.82 \pm 0.03}$. Node removal timescales (red line) remain consistently higher but follow a similar trend. Edge timescales (top right) appear to evolve similarly, with the addition timescale approximated by $\tau_E \propto t^{-0.79 \pm 0.03}$. The ratio of node to edge timescales maintains a stable value of $2.8 \pm 0.3$ throughout the network's evolution.

The IMDb network (bottom panels) exhibits similar patterns but with distinct quantitative differences. Node timescales decrease more slowly, following $\tau_N \propto t^{-0.61 \pm 0.02}$, while edge timescales evolve as $\tau_E \propto t^{-0.58 \pm 0.02}$. The timescale ratio averages $2.3 \pm 0.2$, slightly lower than in the MAG network but equally stable over time. Both addition and removal processes show less volatile year-to-year variations compared to the MAG network, particularly during the interwar period.

These measurements suggest three features that appear to differ from conventional assumptions:
\begin{enumerate}
    \item the timescale separation between node and edge processes remains within a factor of 3, far smaller than the order-of-magnitude separation typically assumed in theoretical models;
    \item both timescales decrease monotonically, indicating continuously accelerating dynamics rather than approach to equilibrium; and,
    \item the stability of timescale ratios suggests a fundamental coupling between node and edge processes that persists despite explosive growth.
\end{enumerate}

\subsection{Node Addition}

\begin{figure*}[htb]
  \centering
  \begin{tabular}{cc}
    \includegraphics[width=0.47\textwidth]{"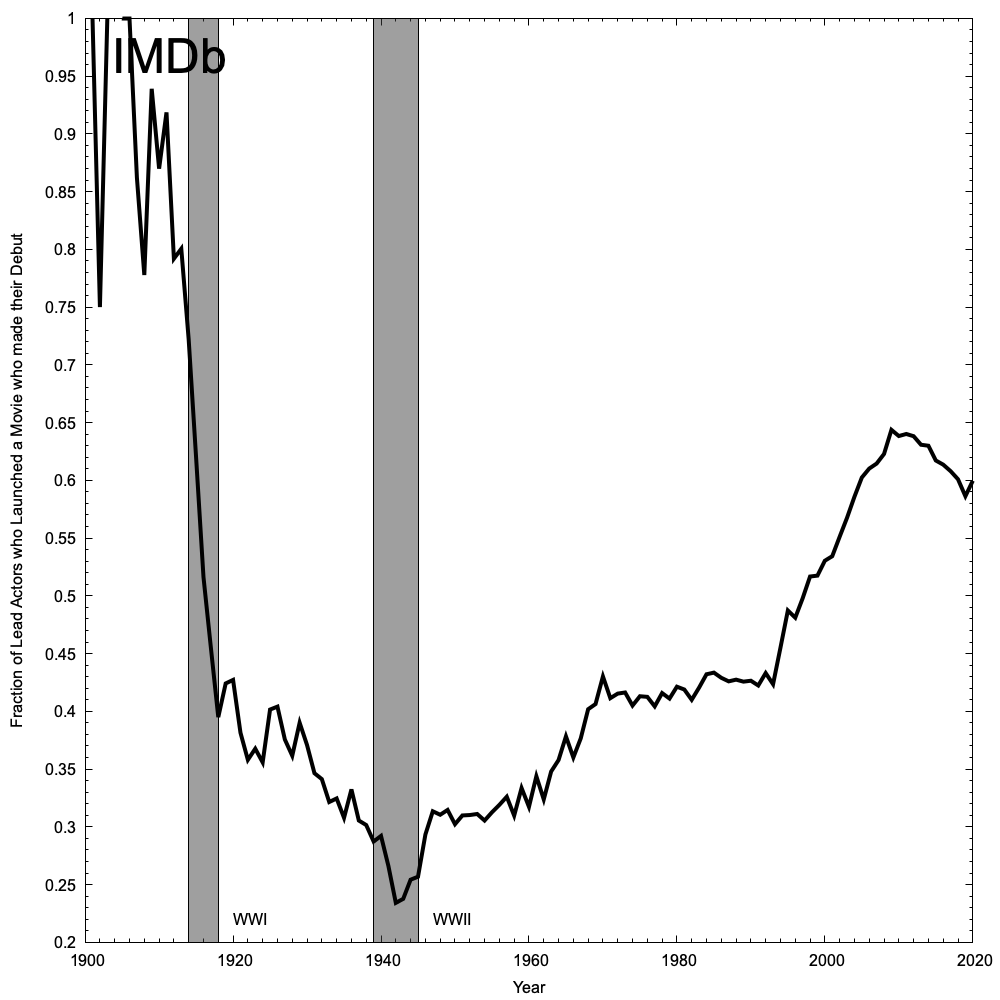"} &
    \includegraphics[width=0.47\textwidth]{"nauthors_new_fraction.png"}
  \end{tabular}
  \caption{The fraction of new participants per year in the MAG (left) and IMDb (right) networks, showing the balance between new entrants and established participants. The academic network shows a gradual decrease in the fraction of new authors over time, while the entertainment network maintains a more stable ratio.}
  \label{node_new_count}
\end{figure*}

The patterns of node addition exhibit distinctive characteristics in the two networks, revealing different sensitivity to external influences. Figure \ref{node_new_count} shows the evolution of new author counts in the MAG network on logarithmic scales, highlighting the power-law regimes and their transition. The pre-1950 growth follows $t^{2.3 \pm 0.1}$, while the post-1950 era shows a steeper $t^{3.1 \pm 0.1}$ acceleration. The transition period (1945-1955) shows heightened volatility, reflecting post-war restructuring of academic research.

Figure \ref{node_new_count} quantifies the balance between new and established participants through the fraction $f_{\text{new}} = N_{\text{new}}/N_{\text{total}}$. In the MAG network (left panel), this fraction shows a systematic decline from 0.90 $\pm$ 0.02 in 1800 to 0.65 $\pm$ 0.02 in 2020, with notable perturbations during major historical events. The World Wars produced sharp drops ($\Delta f \approx -0.1$) followed by compensating increases during recovery periods. The IMDb network (right panel) maintains a more stable fraction around 0.50 $\pm$ 0.03 after 1935, with smaller responses to external events. This contrast suggests fundamental differences in how these professional communities absorb and retain new members.

\subsection{Edge Addition Dynamics}

\begin{figure*}[htb]
  \centering
  \begin{tabular}{cc}
    \includegraphics[width=0.47\textwidth]{"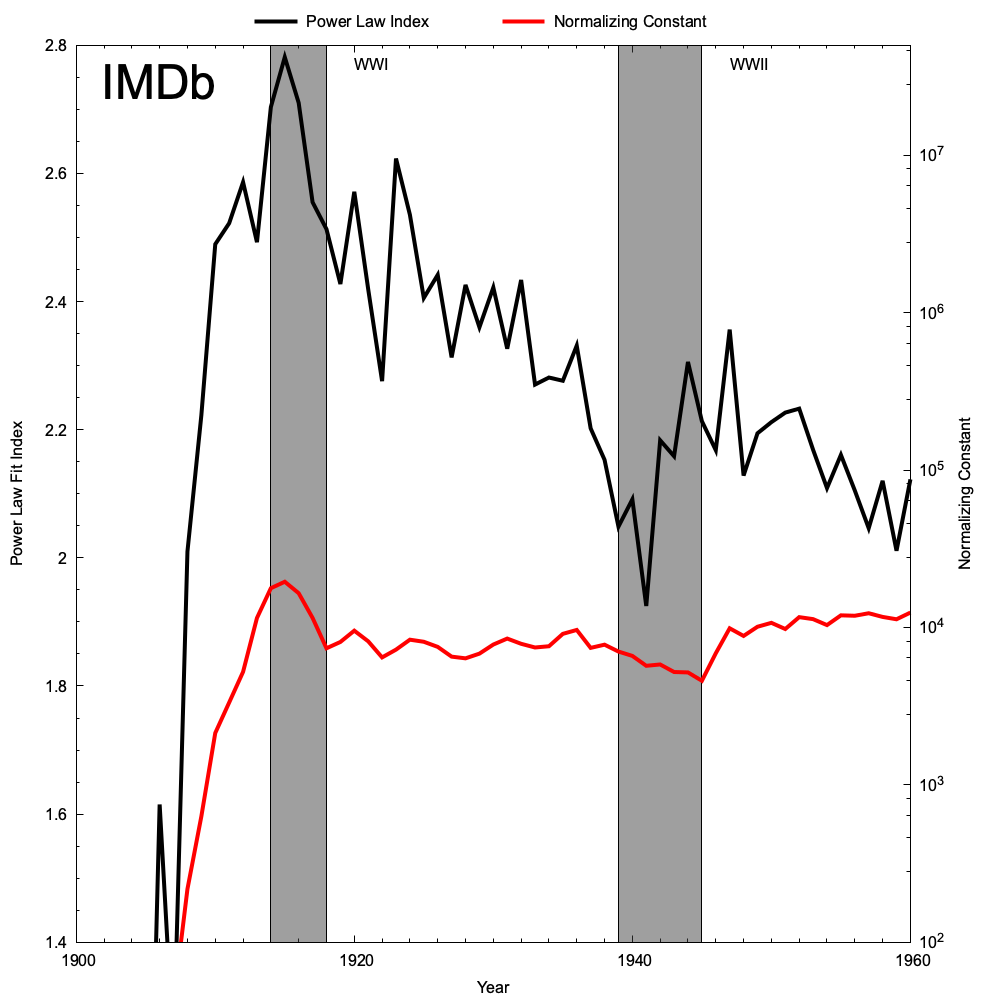"} &
    \includegraphics[width=0.47\textwidth]{"edge-addition-parameters.png"}
  \end{tabular}  
  \caption{Parameter evolution of the power law fits to the edge-addition probability distributions in the MAG and IMDb networks.}
  \label{edges_add_params}
\end{figure*}

\begin{figure*}[htb]
  \centering
  \begin{tabular}{cc}
    \includegraphics[width=0.47\textwidth]{"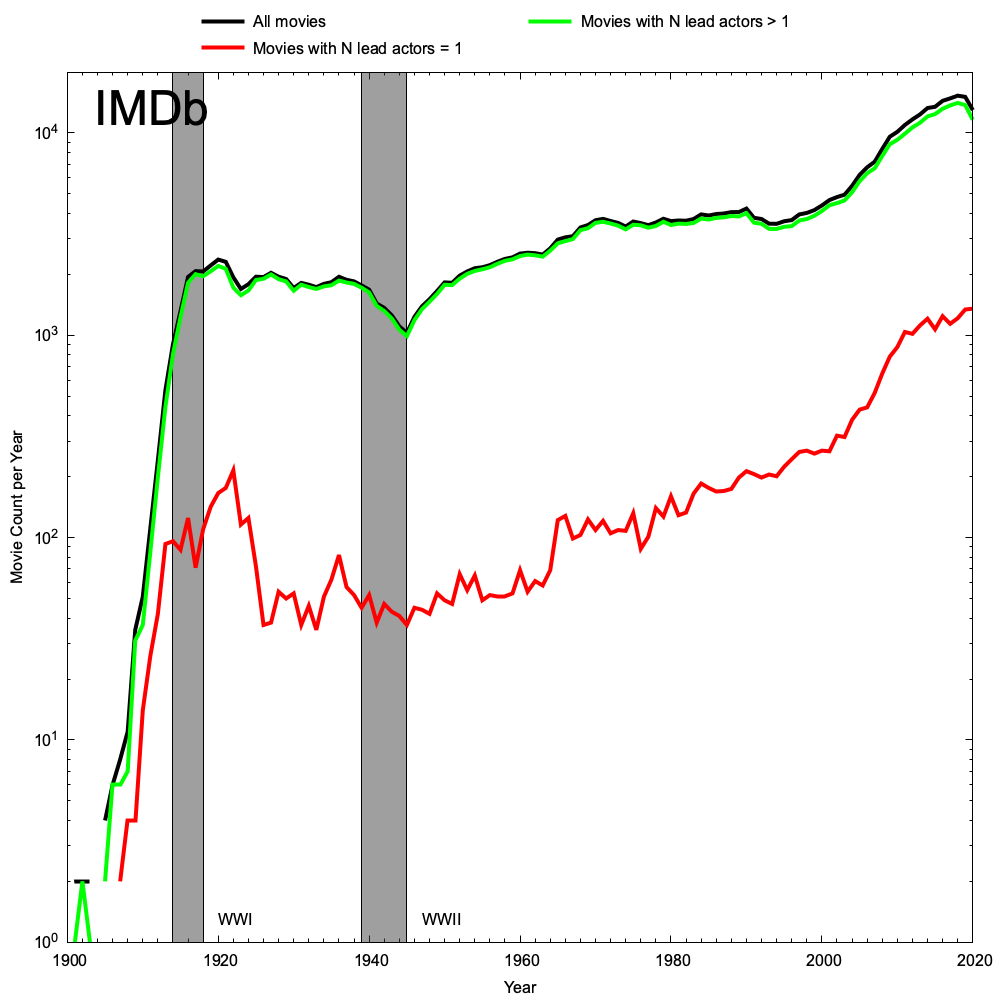"} &
    \includegraphics[width=0.47\textwidth]{"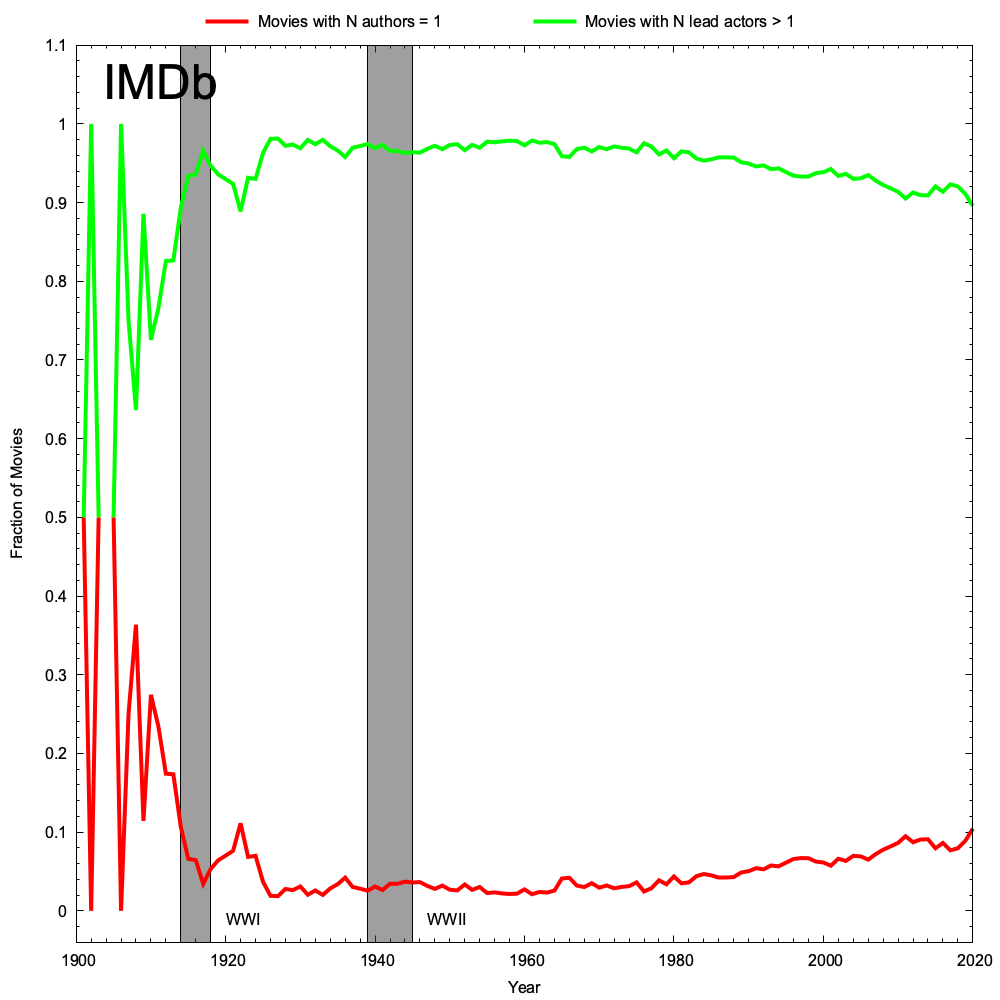"} \\
    \includegraphics[width=0.47\textwidth]{"npapers.png"} &
    \includegraphics[width=0.47\textwidth]{"npapers_fractions.png"}
  \end{tabular}
  \caption{Evolution of collaboration event counts in the networks. Top row shows MAG network: absolute count of papers (left) and relative fractions by author count (right). Bottom row shows IMDb network: absolute count of movies (left) and relative fractions by lead actor count (right). Both networks show systematic changes in the distribution of collaboration size over time.}
  \label{edge_counts}
\end{figure*}

\begin{figure*}[htb]
  \centering
  \begin{tabular}{cc}
    \includegraphics[width=0.47\textwidth]{"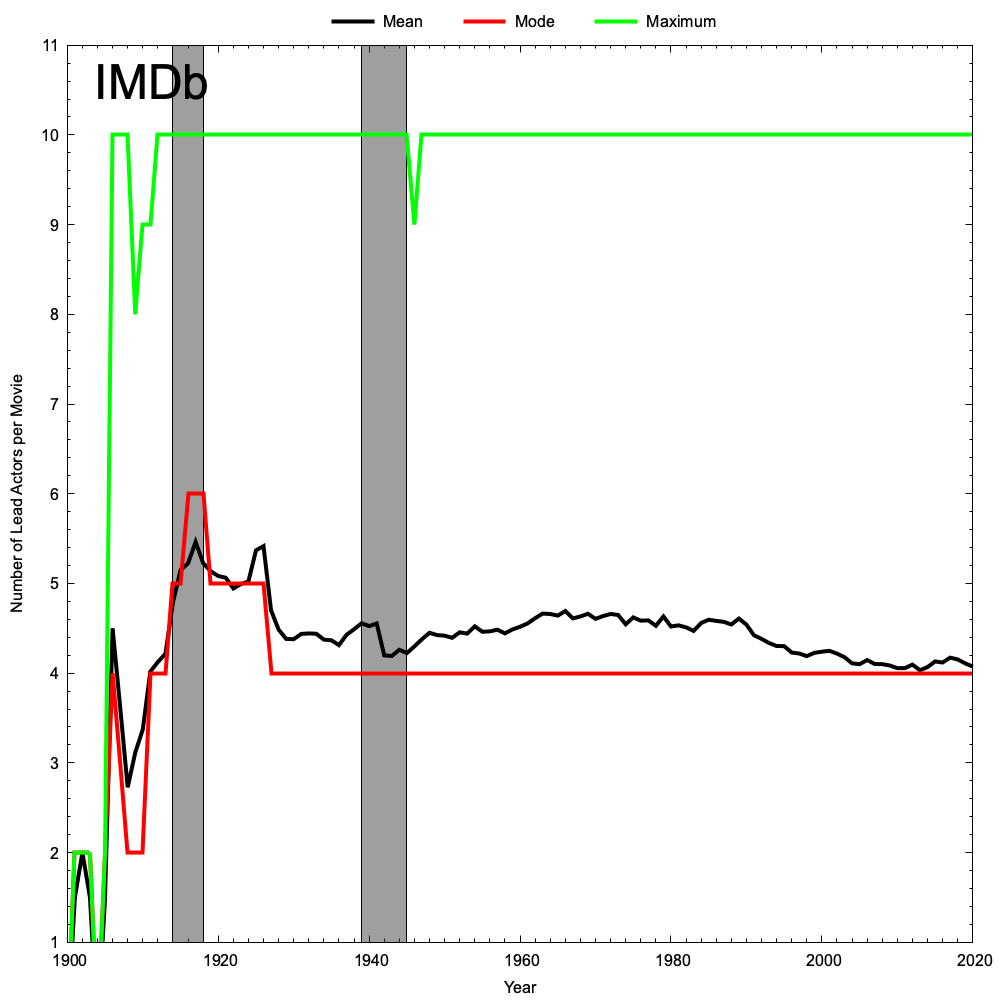"} &
    \includegraphics[width=0.47\textwidth]{"nauthors_per_paper.png"}
  \end{tabular}
  \caption{Statistical measures of collaboration size showing mean, mode, and maximum number of authors per paper (MAG, left) and lead actors per movie (IMDb, right) by year. The evolution of these statistics indicates a trend toward larger collaboration groups in academia, while entertainment industry collaborations remain more stable in size.}
  \label{edge_means}
\end{figure*}

\begin{figure*}[pt]
  \centering
  \begin{tabular}{c}
    \includegraphics[width=0.96\textwidth]{"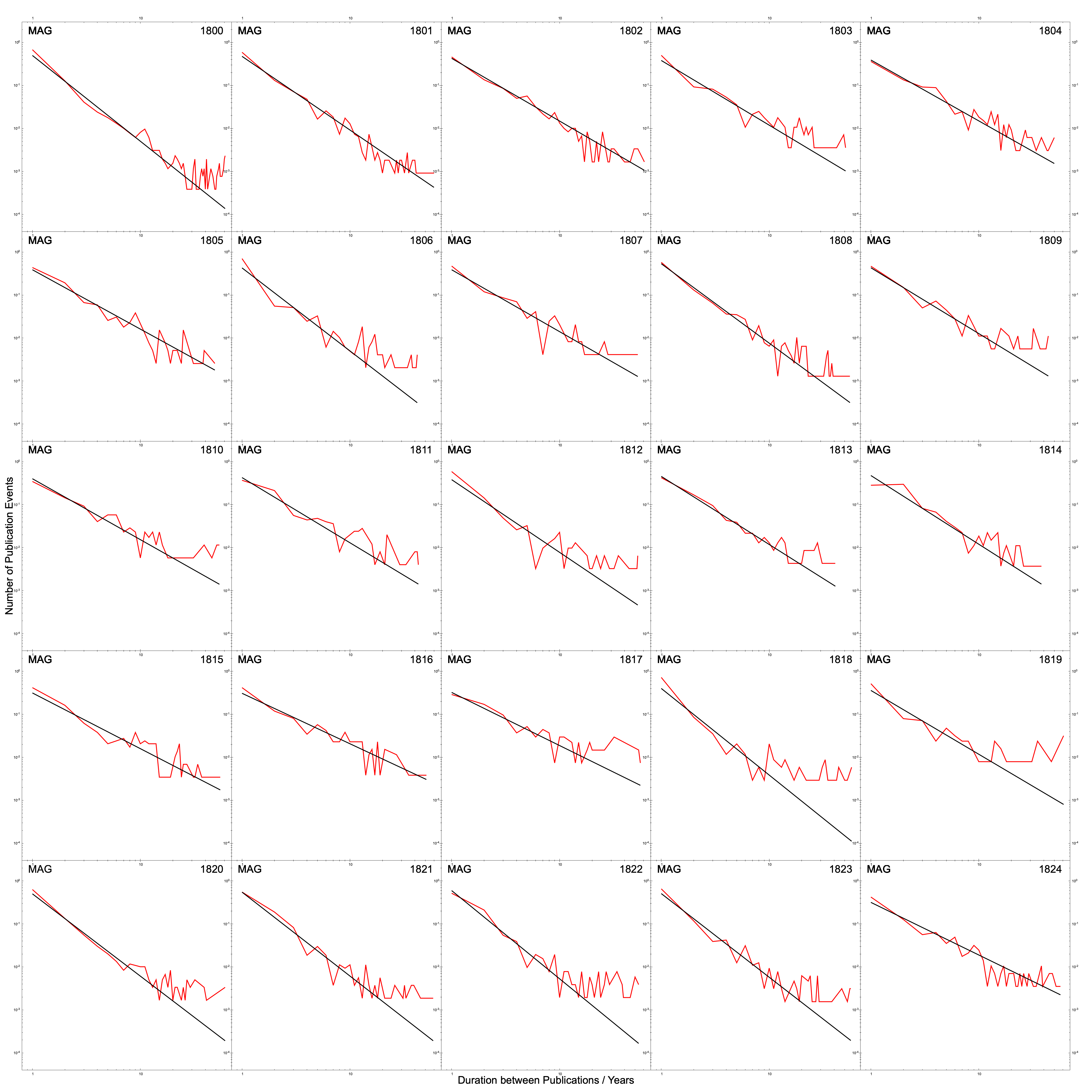"}
  \end{tabular}
  \caption{Probability distribution $P(t)$ of waiting times between successive collaborations, shown for participant cohorts who entered the network between 1800 to 1824. Each point represents the fraction of collaborations that began after a waiting time $t$ (measured in years) following the start of the previous collaboration. The black line shows a power-law fit.}
  \label{mag_edges_add_a}
\end{figure*}

\begin{figure*}[pt]
  \centering
  \begin{tabular}{c}
    \includegraphics[width=0.96\textwidth]{"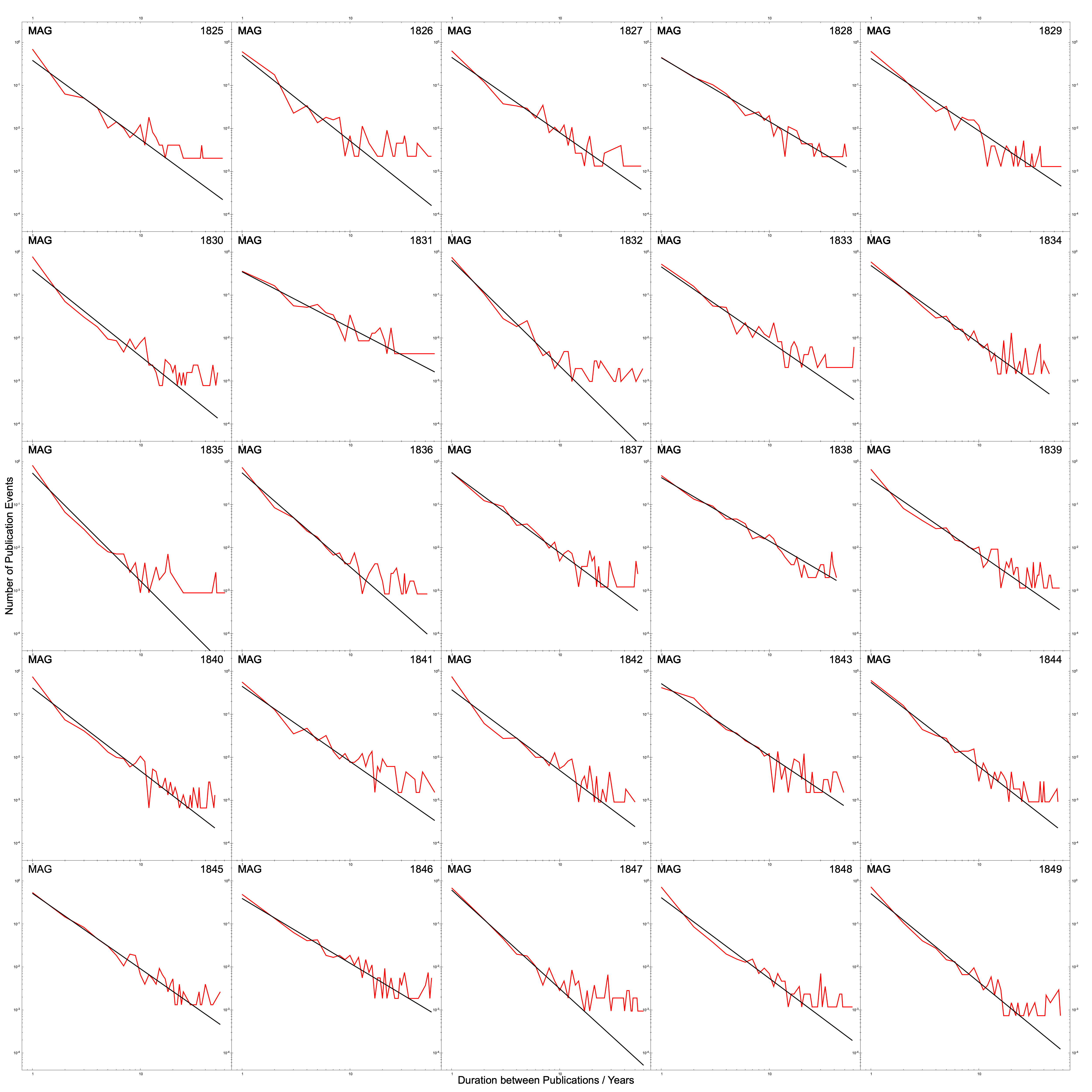"}
  \end{tabular}
  \caption{Probability distribution $P(t)$ of waiting times between successive collaborations, shown for participant cohorts who entered the network between 1825 to 1849. Each point represents the fraction of collaborations that began after a waiting time $t$ (measured in years) following the start of the previous collaboration. The black line shows a power-law fit.}
  \label{mag_edges_add_b}
\end{figure*}

\begin{figure*}[pt]
  \centering
  \begin{tabular}{c}
    \includegraphics[width=0.96\textwidth]{"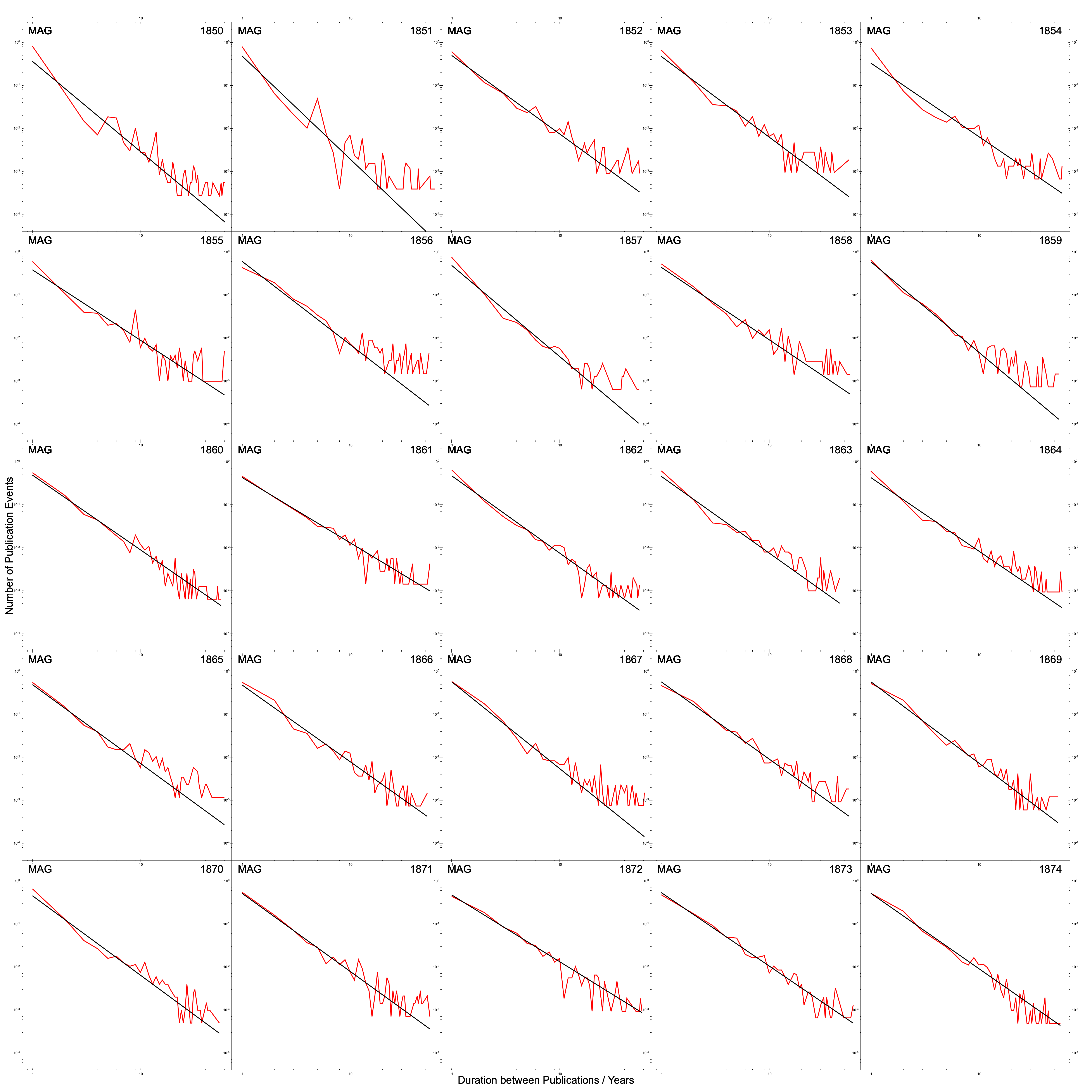"}
  \end{tabular}
  \caption{Probability distribution $P(t)$ of waiting times between successive collaborations, shown for participant cohorts who entered the network between 1850 to 1874. Each point represents the fraction of collaborations that began after a waiting time $t$ (measured in years) following the start of the previous collaboration. The black line shows a power-law fit.}  
  \label{mag_edges_add_c}
\end{figure*}

\begin{figure*}[pt]
  \centering
  \begin{tabular}{c}
    \includegraphics[width=0.96\textwidth]{"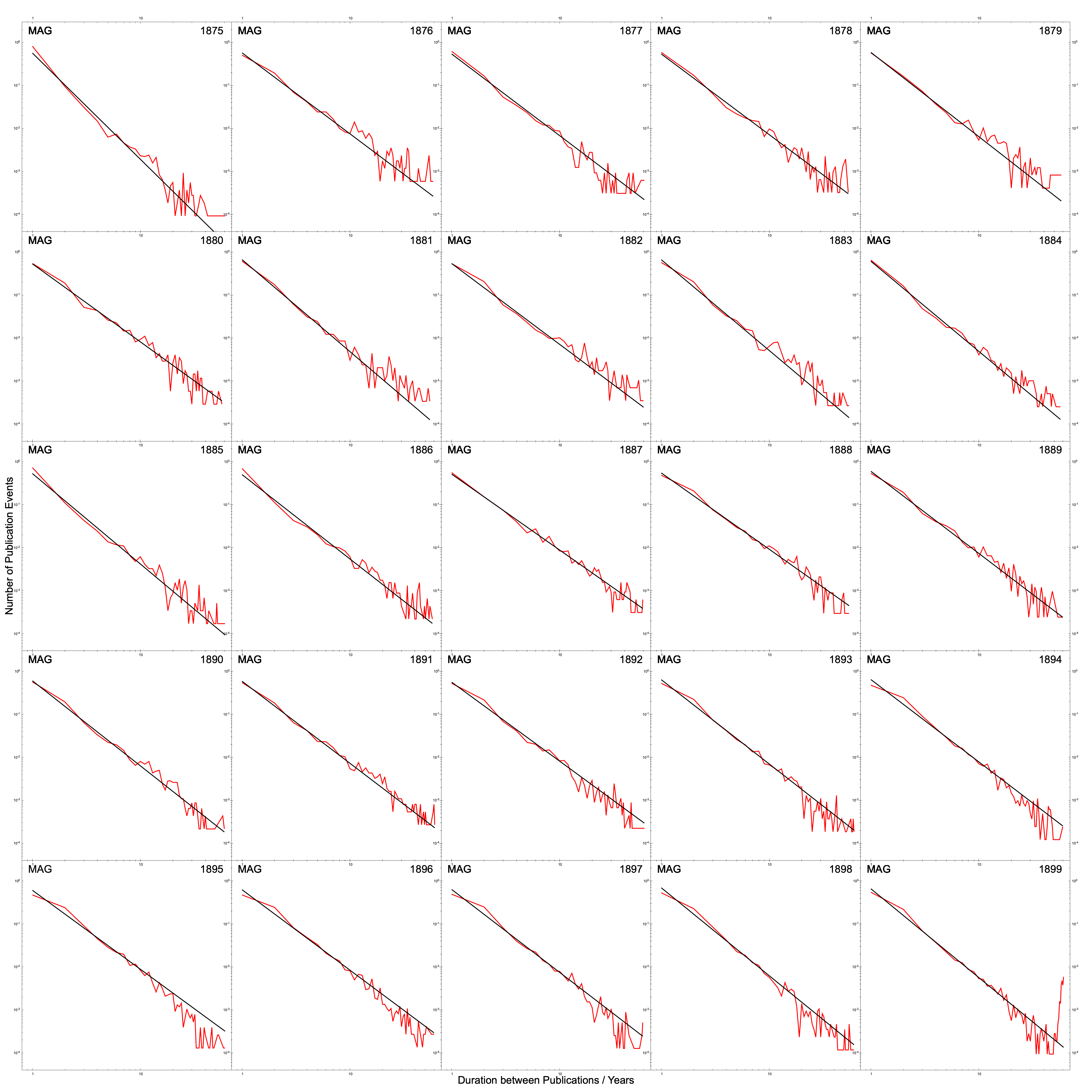"}
  \end{tabular}
  \caption{Probability distribution $P(t)$ of waiting times between successive collaborations, shown for participant cohorts who entered the network between 1875 to 1899. Each point represents the fraction of collaborations that began after a waiting time $t$ (measured in years) following the start of the previous collaboration. The black line shows a power-law fit.}
  \label{mag_edges_add_d}
\end{figure*}

\begin{figure*}[pt]
  \centering
  \begin{tabular}{c}
    \includegraphics[width=0.96\textwidth]{"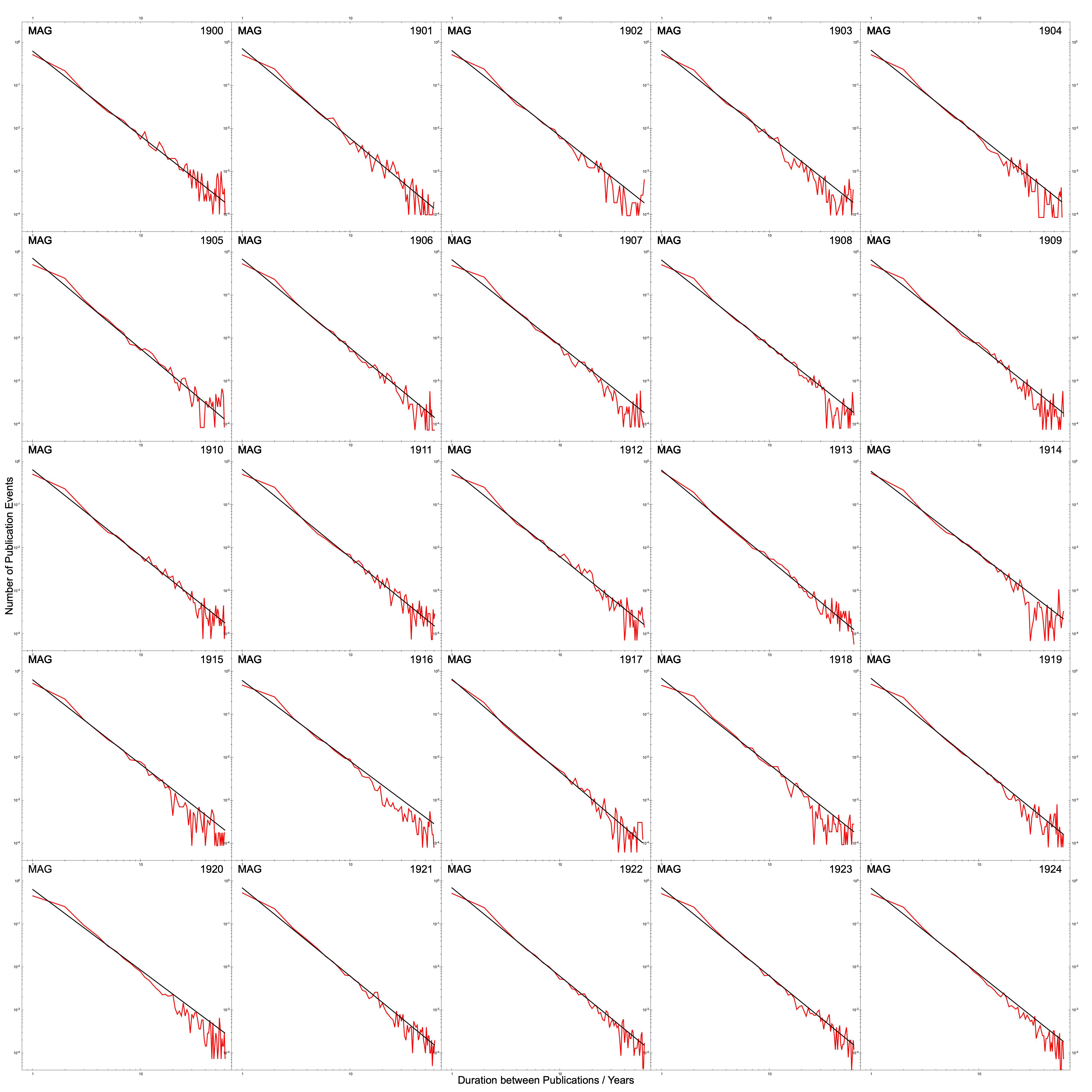"}
  \end{tabular}
  \caption{Probability distribution $P(t)$ of waiting times between successive collaborations, shown for participant cohorts who entered the network between 1900 to 1924. Each point represents the fraction of collaborations that began after a waiting time $t$ (measured in years) following the start of the previous collaboration. The black line shows a power-law fit.}
  \label{mag_edges_add_e}
\end{figure*}

\begin{figure*}[pt]
  \centering
  \begin{tabular}{c}
    \includegraphics[width=0.96\textwidth]{"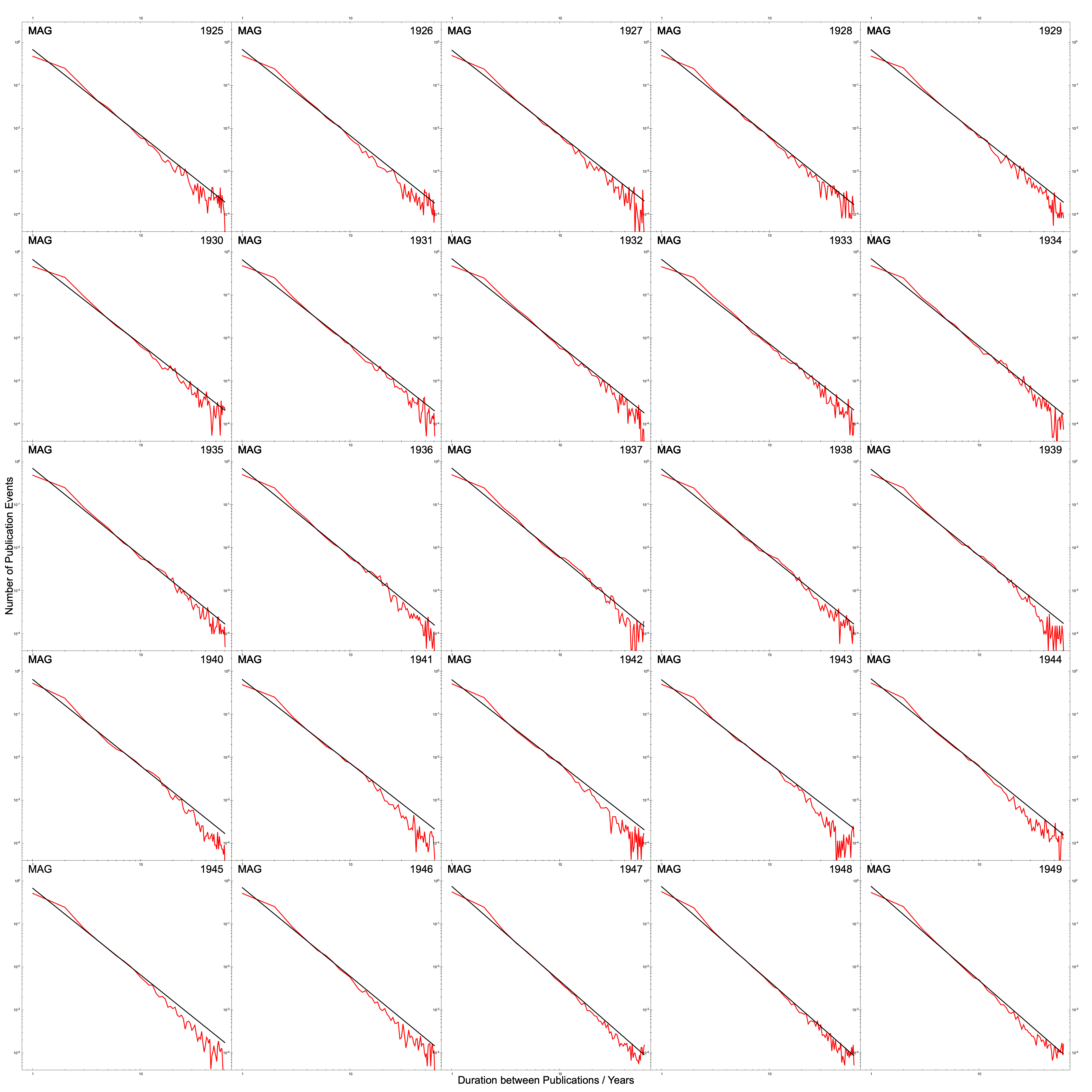"}
  \end{tabular}
  \caption{Probability distribution $P(t)$ of waiting times between successive collaborations, shown for participant cohorts who entered the network between 1925 to 1949. Each point represents the fraction of collaborations that began after a waiting time $t$ (measured in years) following the start of the previous collaboration. The black line shows a power-law fit.}
  \label{mag_edges_add_f}
\end{figure*}

\begin{figure*}[pt]
  \centering
  \begin{tabular}{c}
    \includegraphics[width=0.96\textwidth]{"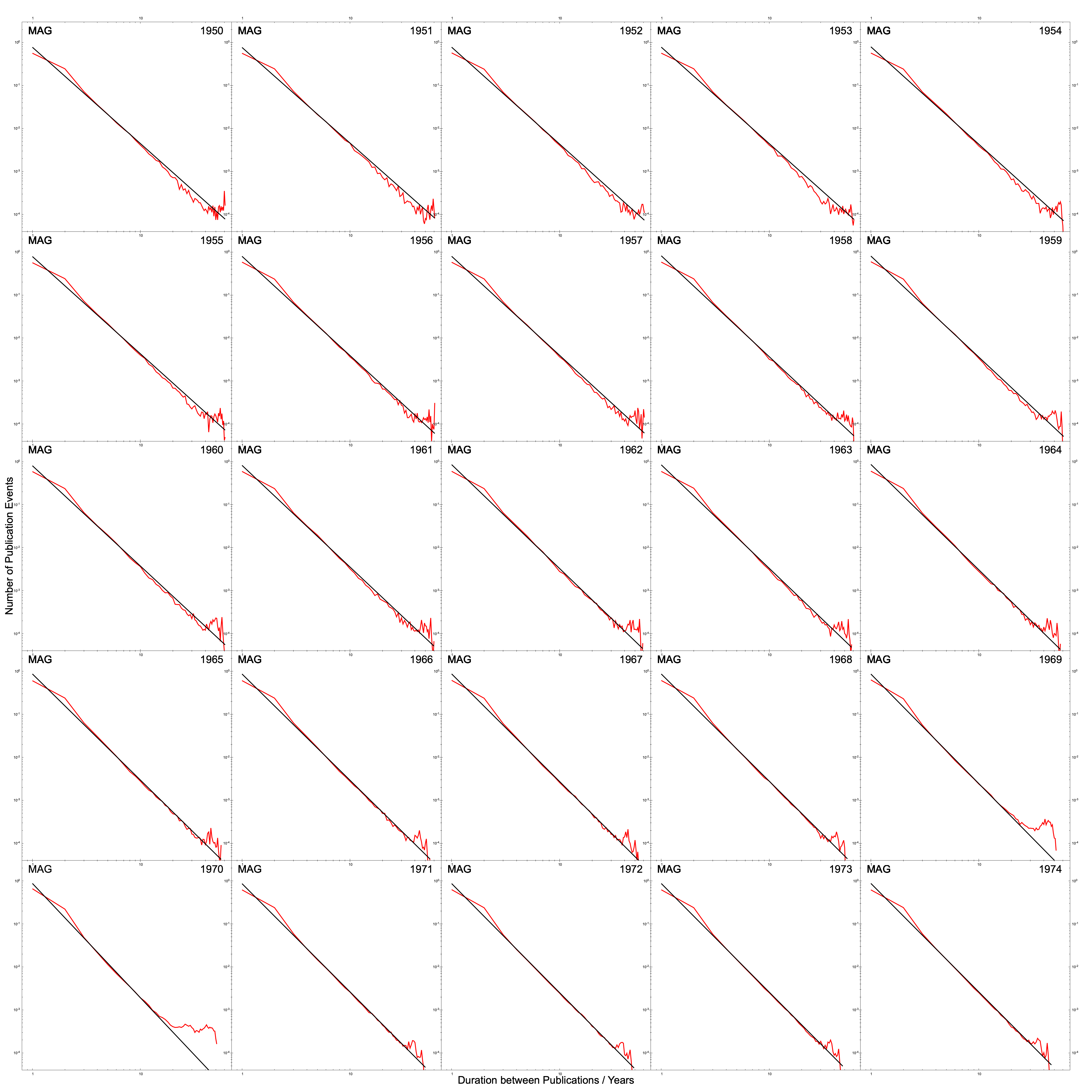"}
  \end{tabular}
  \caption{Probability distribution $P(t)$ of waiting times between successive collaborations, shown for participant cohorts who entered the network between 1950 to 1974. Each point represents the fraction of collaborations that began after a waiting time $t$ (measured in years) following the start of the previous collaboration. The black line shows a power-law fit.}
  \label{mag_edges_add_g}
\end{figure*}

\begin{figure*}[pt]
  \centering
  \begin{tabular}{c}
    \includegraphics[width=0.96\textwidth]{"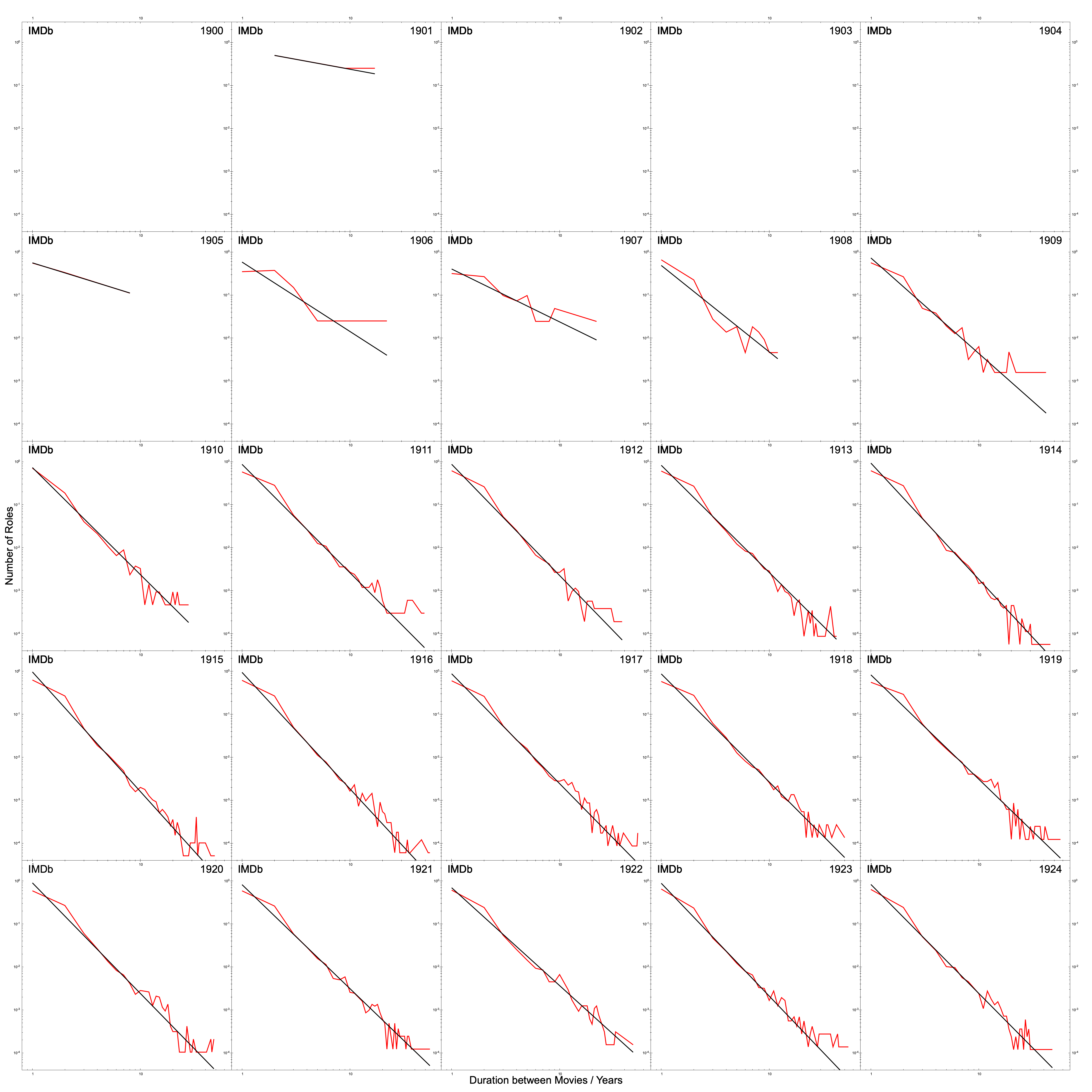"}
  \end{tabular}
  \caption{Probability distribution $P(t)$ of waiting times between successive collaborations, shown for participant cohorts who entered the network between 1900 to 1924. Each point represents the fraction of collaborations that began after a waiting time $t$ (measured in years) following the start of the previous collaboration. The black line shows a power-law fit.}
  \label{imdb_edges_add_a}
\end{figure*}

\begin{figure*}[pt]
  \centering
  \begin{tabular}{c}
    \includegraphics[width=0.96\textwidth]{"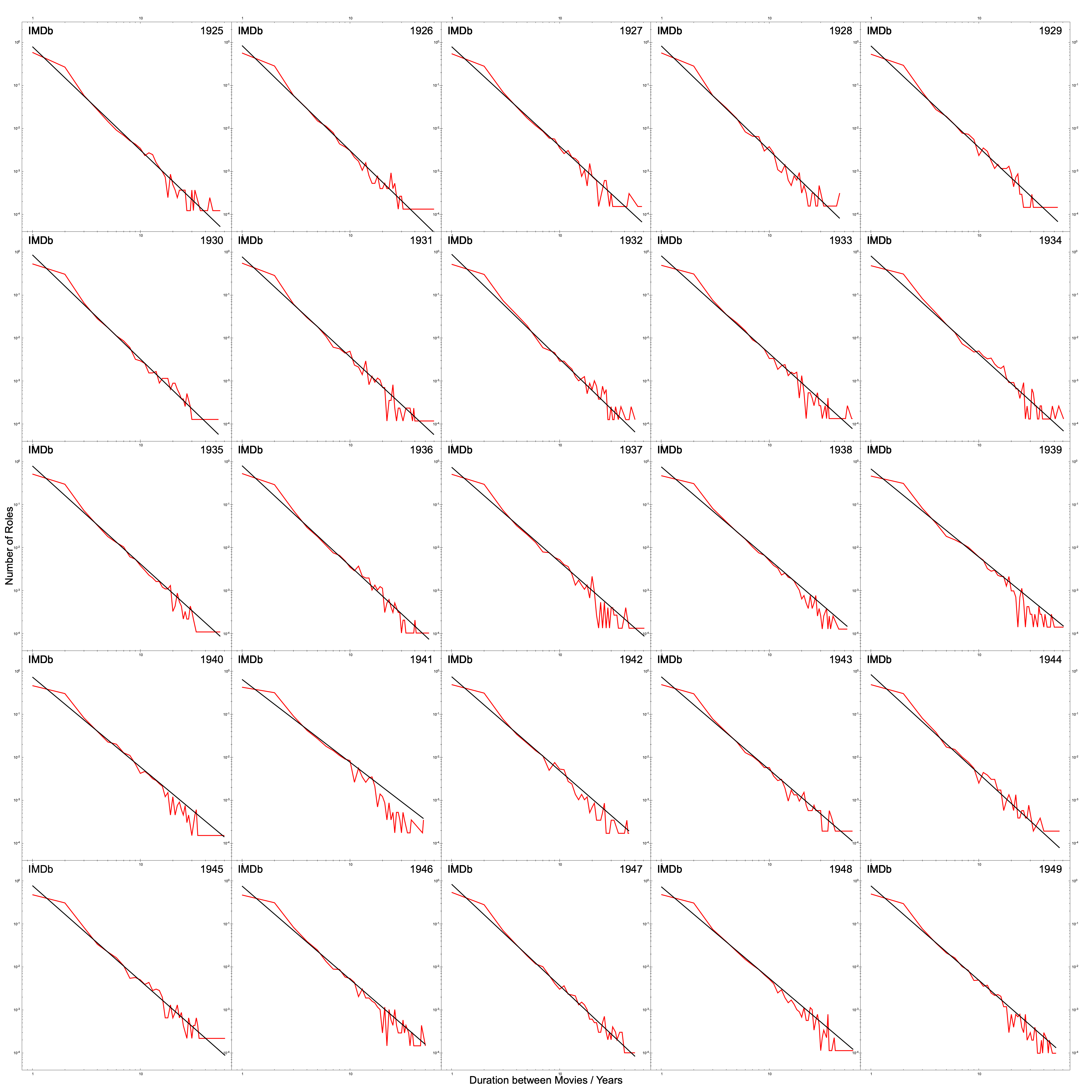"}
  \end{tabular}
  \caption{Probability distribution $P(t)$ of waiting times between successive collaborations, shown for participant cohorts who entered the network between 1925 to 1949. Each point represents the fraction of collaborations that began after a waiting time $t$ (measured in years) following the start of the previous collaboration. The black line shows a power-law fit.}
  \label{imdb_edges_add_b}
\end{figure*}

\begin{figure*}[pt]
  \centering
  \begin{tabular}{c}
    \includegraphics[width=0.96\textwidth]{"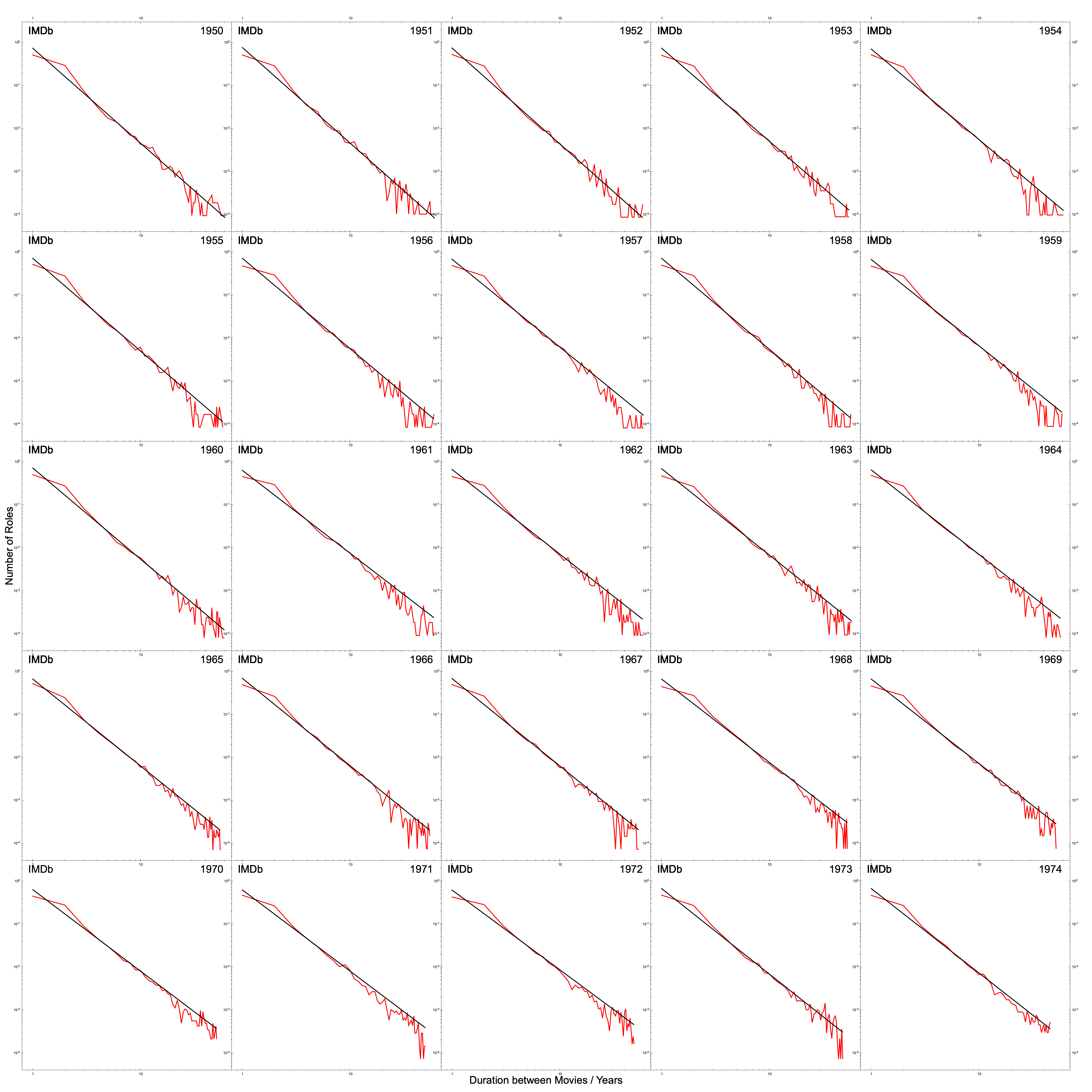"}
  \end{tabular}
  \caption{Probability distribution $P(t)$ of waiting times between successive collaborations, shown for participant cohorts who entered the network between 1950 to 1974. Each point represents the fraction of collaborations that began after a waiting time $t$ (measured in years) following the start of the previous collaboration. The black line shows a power-law fit.}
  \label{imdb_edges_add_c}
\end{figure*}

The temporal patterns of collaborative behaviour are revealed through analysis of ``waiting times'' between successive collaborations, as illustrated in Figures \ref{mag_edges_add_a} through \ref{mag_edges_add_g} for the Microsoft Academic Graph (MAG) and Figures \ref{imdb_edges_add_a} through \ref{imdb_edges_add_c} for the Internet Movie Database (IMDb). For each participant in these networks, the waiting time $t$ is defined as the interval between the start of one project and the start of the next. The probability distribution of these waiting times, $P(t)$, provides fundamental insights into the rhythm and patterns of professional collaborative activity.

To examine how waiting time patterns evolved over the network's history, we conducted a cohort-based analysis of these distributions. For each year, all nodes (authors or actors) who first joined the network during that year were identified, forming a temporal cohort. The waiting time distribution $P(t)$ was then computed separately for each cohort by analysing all subsequent collaborations of its members. Figures \ref{mag_edges_add_a} through \ref{imdb_edges_add_c} present these cohort-specific distributions, with each panel showing $P(t)$ for a different entry year. This approach allows us to distinguish between historical changes in collaborative behaviour and effects that might arise from mixing different generations of participants.

These distributions reveal a broad spectrum of collaborative behaviours. At one end, short waiting times ($t < 1$ year) correspond to periods of intense activity, where participants transition rapidly between successive projects. At the other extreme, extended waiting times ($t \gg 1$ year) indicate prolonged gaps in collaborative activity, which may represent temporary withdrawals from the field or shifts in professional focus. The observed power-law form of these distributions, $P(t) \sim t^{-\alpha}$, suggests that collaborative activity lacks a characteristic timescale. Instead, waiting times follow a scale-free pattern, spanning from months to decades, with the probability of longer gaps decreasing according to a universal scaling relationship.

The Microsoft Academic Graph (MAG) network exhibits a systematic evolution in its waiting time distributions over the past two centuries. During the early period (1800-1850), the distributions follow steep power laws with indices of $\gamma \approx 2.3 \pm 0.2$, indicating that authors who remained active in publishing typically maintained short intervals between successive publications. The probability of extended gaps between publications decreased rapidly with time, following the power-law relationship $P(t) \sim t^{-\gamma}$. This pattern underwent a gradual but significant transformation over subsequent decades. By 1900, the power-law index had decreased to $\gamma \approx 1.9 \pm 0.1$, continuing its decline to reach $\gamma \approx 1.6 \pm 0.1$ by 1950. This progressive flattening of the distributions suggests a fundamental shift in academic working patterns, with longer intervals between publications becoming increasingly common. Such evolution may reflect the growing complexity and resource requirements of research projects over time.

The Internet Movie Database (IMDb) network, illustrated in Figures \ref{imdb_edges_add_a} through \ref{imdb_edges_add_c}, reveals distinct temporal patterns in professional collaboration. The waiting time distributions tend to show steeper power-law behavior compared to the academic network, with indices decreasing from $\gamma \approx 2.6 \pm 0.1$ in the earliest cohorts to $\gamma \approx 2.1 \pm 0.1$ by 1950. These steeper distributions reflect the more structured nature of professional patterns in the entertainment industry, where regular engagement in projects represents the norm and extended gaps between collaborations occur less frequently. While the declining power-law index indicates a gradual diversification of career patterns over time, the entertainment industry maintained more consistent collaborative rhythms compared to academic research throughout the studied period.

The data suggests a deviation from pure power-law behaviour in both networks at very short time intervals ($t < 6$ months). This pattern of fewer extremely short waiting times suggests characteristic minimum timescales inherent to collaborative processes in each domain. In academia, this minimum threshold reflects the practical constraints of conducting research, analysing results, and preparing manuscripts for publication. Similarly, in the entertainment industry, this threshold corresponds to the minimum time required for movie production, from filming through post-production. The temporal evolution of these power-law indices, visualised in Figure \ref{edges_add_params}, illuminates how collaborative patterns have developed along distinct trajectories in academic and entertainment contexts, while maintaining these fundamental minimum timescales.

The notable consistency of these waiting time distributions' functional form, persisting across historical periods that witnessed dramatic changes in both network size and collaboration frequency, points to deep-seated patterns in human collaborative behaviour. While the frequency of collaboration has grown over time in both domains, the relative distribution of waiting times between successive collaborations has maintained its characteristic shape within each professional sphere. This invariance suggests that despite the vast expansion in the scale and scope of collaborative enterprises, the fundamental temporal rhythms of professional activity remain governed by consistent patterns. These patterns may reflect intrinsic aspects of how humans organise and pace their collaborative engagements, transcending the substantial changes in technology, institutional structures, and professional practices that have occurred over the studied period.

\subsection{Integrated Growth Dynamics}

The temporal evolution of collaboration size distributions is presented in Figure \ref{edge_counts}, with absolute counts shown in the left panels and relative proportions in the right panels. The Microsoft Academic Graph (MAG) network shows a dramatic transformation in authorship patterns over the past century. Prior to 1920, single-author papers constituted the overwhelming majority ($>80\%$) of academic publications. Subsequently, a systematic shift toward multi-author collaborations emerged. Publications with moderate team sizes ($1 < n < 7$ authors) exhibited strong growth, eventually comprising more than half of all papers by 1980. Large-team collaborations ($n > 6$ authors) also gained prominence, representing approximately $10\%$ of publications by 2020. In contrast, the Internet Movie Database (IMDb) network maintained relatively stable collaboration size distributions throughout its history, although a subtle trend toward larger cast sizes became apparent in recent decades.

Statistical measures of these collaboration patterns are shown in Figure \ref{edge_means}. The academic sphere, represented by the MAG network (left panel), shows a gradual increase in average collaboration size, from $1.2 \pm 0.1$ authors per paper in 1800 to $5.8 \pm 0.2$ by 2020. A notable change occurred around 1995 when the modal collaboration size shifted from single authorship to three authors. The maximum team size has increased over time, with some contemporary publications involving more than 100 authors. The entertainment industry, as captured by the IMDb network (right panel), exhibits quite different dynamics. Mean cast sizes increased modestly from $3.2 \pm 0.2$ to $4.5 \pm 0.2$ between 1900 and 2020. While maximum cast sizes exhibit periodic fluctuations, they show no sustained long-term growth, suggesting persistent production constraints in the film industry.

These quantitative measures reveal different evolutionary patterns in academic and entertainment collaborations. The academic sphere has experienced a consistent trend toward larger research teams, with this tendency accelerating after 1950. This change may reflect increasing research complexity, technological advancement, and a growing emphasis on interdisciplinary approaches. In contrast, the entertainment industry has maintained relatively stable collaboration size distributions, indicating distinct professional constraints and incentive structures that have remained largely unchanged despite technological and cultural evolution.

\section{Discussion}

The quantitative evidence from both networks suggests patterns of collaborative network evolution that may require extensions to existing theoretical frameworks. These patterns are characterised by several interconnected features:
\begin{enumerate}
\item \textbf{Unprecedented Growth Dynamics:} Both networks show persistent super-linear growth that shows no indication of saturation. The Microsoft Academic Graph exhibits two distinct power-law regimes, transitioning from $t^{2.3}$ before 1950 to an accelerated $t^{3.1}$ afterward. The Internet Movie Database maintains consistent $t^{1.8}$ growth throughout its evolution. These growth rates exceed both demographic expansion and technological scaling factors.

\item \textbf{Synchronised Network Processes:} The temporal evolution of nodes and edges maintains stable relationships, characterised by constant timescale ratios ($2.8 \pm 0.3$ for academic collaboration, $2.3 \pm 0.2$ for entertainment networks). These coupled processes follow parallel power-law trajectories, and have a synchronised acceleration across all aspects of network development.

\item \textbf{External Environmental Influence:} Network dynamics are sensitive to major historical events, evident in academic collaboration where World Wars induced substantial perturbations ($10$-$15\%$) in the rate of new participant entry. However, edge formation processes and internal network characteristics, such as collaboration size distributions, exhibited greater resilience to these external disruptions.

\item \textbf{Evolution of Collaborative Rhythms:} Both networks display characteristic power-law distributions in inter-collaboration waiting times, though with distinct evolutionary trajectories. Academic collaboration patterns show systematic flattening over time (power-law index decreasing from $\gamma \approx 2.3$ to $\gamma \approx 1.6$), while entertainment industry patterns maintain steeper distributions (evolving from $\gamma \approx 2.6$ to $\gamma \approx 2.1$).

\item \textbf{Contrasting Collaboration Dynamics:} Academic research shows a systematic evolution toward larger collaborative teams, with mean authorship increasing from $1.2$ to $5.8$ participants and maximum team sizes exceeding $100$ members. In contrast, entertainment industry collaborations maintain stable size distributions (mean cast size rising from $3.2$ to $4.5$) with well-defined upper bounds, indicating distinct underlying constraints and incentives.
\end{enumerate}
These empirical observations suggest a potential need to reconsider three common assumptions:
\begin{itemize}
\item the prevailing expectation of asymptotic or saturating network growth,
\item the hypothesised independence of node and edge evolutionary timescales, and
\item the conventional treatment of collaborative networks as closed systems isolated from external influences.
\end{itemize}

The observed growth patterns in these collaboration networks appear to differ from some common assumptions in network science. While some features align with fitness-based preferential attachment models \citep{bianconi2001}, the sustained acceleration over centuries indicates additional mechanisms must be at work. Traditional network models, from the classic Barabási-Albert preferential attachment \citep{barabasi1999} to more recent fitness-based approaches \citep{kong2019}, typically assume linear or sub-linear growth rates. These assumptions arise partly from computational convenience and partly from the expectation that social networks should exhibit saturation. However, our findings indicate that collaboration networks may follow different patterns, with sustained growth rates and power-law exponents above unity.

This observation might appear to contradict several influential studies. \citet{newman2001} reported linear growth in scientific collaboration networks, while \citet{barabasi2002} proposed widely adopted preferential attachment models with linear growth. However, these studies analysed relatively short time periods, typically five years or fewer, within which the acceleration we observe might appear approximately linear. More recent work has begun to hint at these dynamics, with \citet{dong2017} noting accelerating collaboration rates in physics networks from 1985 to 2015, though without quantifying the power-law nature of this growth. \citet{zhang2023} suggested universal patterns in network growth, albeit over shorter timescales. Our unprecedented 220-year temporal span reveals the limitations of extrapolating from shorter-term observations, suggesting these apparent contradictions reflect differences in scale rather than fundamental disagreement. This interpretation aligns with \citet{liu2021}'s evidence for external event influence on network evolution, though their analysis covered smaller scales and shorter durations.

The closest theoretical parallel to our observations may be the densification power laws described by \citet{leskovec2008}, though our data show even stronger acceleration over longer periods. However, biological growth processes might offer more relevant models than traditional network frameworks. While individual organisms typically follow sigmoidal growth curves that saturate at maturity \citep{west2001}, larger-scale biological systems can exhibit sustained acceleration. The evolution of multicellularity shows how hierarchical organisation enables continuous super-linear scaling \cite{bonner2004}, while neural network development during brain growth shows phases of explosive connectivity increase \citep{johnson2003}, driven by both intrinsic programs and environmental interactions.

Particularly relevant is the metabolic scaling theory developed by \citet{west1997}, which shows how power-law relationships naturally emerge in biological networks optimised for resource distribution, with exponents similar to those observed in our collaboration networks. This biological perspective helps explain our observation of stable local properties (waiting time distributions) amid global acceleration, mirroring how biological systems maintain homeostasis at small scales while enabling growth at larger scales \citep{kleiber1947}. Such biological analogies suggest collaboration networks might be better understood through models of organic growth and development, where local constraints coexist with system-wide acceleration.

The observed relationship between node and edge timescales indicates potential refinements to established network theory. Conventional network models typically assume a clear separation between these timescales, with edge dynamics operating much faster than node turnover \citep{holme2019temporal}. However, our analysis reveals that these processes maintain a stable albeit noisy coupling ratio ($\tau_N/\tau_E \approx 2.8 \pm 0.3$) over centuries of network evolution. This persistent coupling suggests that the traditional assumption of timescale separation, while mathematically convenient, may have led to oversimplified models that fail to capture essential features of real-world network dynamics.

This finding may have implications for two major theoretical frameworks in network science. First, it suggests potential refinements to adaptive network models \citep{gross2008}, which have typically assumed that network topology evolves much more rapidly than node composition. Second, it indicates opportunities to extend temporal network theory. As \citet{holme2012} detailed in their comprehensive review, the assumption of separated timescales has been an important principle in analysing network temporal evolution since the seminal work of \citet{barabasi1999}. The observation that node and edge processes appear to maintain consistent coupling over long time periods suggests that these theoretical frameworks might benefit from additional development, particularly when addressing the long-term evolution of social and collaborative networks.

The persistent coupling between these timescales indicates that collaborative networks maintain a kind of dynamic equilibrium, where the processes of adding new participants and forming new connections remain fundamentally interlinked. This relationship may reflect deeper organisational principles in how human collaborative systems develop and adapt over time. Understanding the mechanisms that maintain this stable coupling ratio could provide crucial insights for developing more accurate models of network evolution.

The significant impact of external events on network dynamics, particularly evident in the Microsoft Academic Graph (MAG), reveals fundamental limitations in treating collaborative networks as isolated systems. Our long-term analysis suggests that external influences shape network evolution in profound and lasting ways. During the World Wars, for instance, the rate of new node addition in academic networks showed substantial perturbations of $10-15\%$ from baseline growth trends. These effects persisted well beyond the duration of the events themselves, suggesting complex mechanisms of network response and recovery.

This environmental coupling manifests differently across network processes. Node dynamics—the entry of new participants into the network—show strong sensitivity to external events. During periods of global disruption, the fraction of new authors entering the academic network show clear deviations from long-term trends. However, edge formation processes and internal network statistics, such as collaboration size distributions, exhibit greater resilience. This differential response suggests a hierarchical structure in network sensitivity, where the recruitment of new participants proves more vulnerable to external perturbation than the maintenance of existing collaborative relationships.

Previous research by \citet{karimi2018} identified certain external influences on network evolution, but their analysis, limited to shorter time periods, characterised these as temporary perturbations to an otherwise autonomous system. Our multi-century analysis reveals a different picture: external events act as persistent shapers of network evolution, leaving lasting signatures on network structure and dynamics. These signatures appear not only in immediate network responses but also in long-term adaptive changes to collaboration patterns.

The implications of this environmental coupling extend beyond simple cause-and-effect relationships. Historical events can trigger cascading changes in network dynamics, altering both the rate and pattern of network growth. For example, post-war periods often show not just recovery to previous growth patterns but qualitative shifts in collaboration dynamics, suggesting that external perturbations can drive the network into new evolutionary regimes. This observation aligns with complex systems theory, where external forcing can induce phase transitions in system behaviour.

These findings suggest that realistic models of network evolution must incorporate sophisticated mechanisms of environmental coupling. Rather than treating external events as noise or perturbation, models should account for how networks respond to and are shaped by their historical context. This perspective aligns with recent work in adaptive networks but requires extending these frameworks to account for long-term environmental influences on network evolution.

The analysis of collaboration waiting time distributions reveals a striking pattern of behavioural stability amid accelerating network growth. These distributions, which measure the intervals between successive collaborations, follow power laws with evolving indices. In the academic network, these indices decrease from $\gamma \approx 2.3$ to $1.6$, while the entertainment network shows a more modest decline from $\gamma \approx 2.6$ to $2.1$. The persistence of these power-law forms, despite dramatic network expansion, suggests fundamental constraints on human collaborative behaviour. The steeper distributions in entertainment ($\gamma \approx 2.1$ versus $1.6$ in academia by 1950) reflect distinct professional rhythms: entertainment careers demand more frequent collaboration, while academic work permits longer intervals between projects. This finding extends previous research on human dynamics \citep{karsai2011} to multi-century timescales, implying that such behavioural patterns can remain stable even as the surrounding system undergoes dramatic growth.

The distinct evolutionary trajectories of academic and entertainment networks illuminate how institutional and social structures shape network development. The academic network appears to show greater environmental sensitivityand more pronounced acceleration, while the entertainment network maintains more stable growth patterns. These differences reflect their underlying organisational frameworks: academic collaboration has been shaped by increasing institutionalisation and formal research structures, while entertainment industry growth appears constrained by market forces and production requirements. A detailed examination of these institutional influences will be presented in a subsequent paper in this series.

These findings suggest several directions for future research.

First, new theoretical frameworks are needed that can accommodate explosive growth while maintaining stable local properties. Such models might incorporate hierarchical structure \citep{ravasz2003} or multi-scale dynamics \citep{benson2018} to capture the observed combination of global acceleration and local stability.

Second, the strong coupling between external events and network evolution calls for the development of coupled dynamical systems approaches. These might combine network evolution with models of institutional growth, technological advancement, or socioeconomic development.

Third, the similarity of node and edge timescales suggests the need for new analytical techniques that do not rely on timescale separation. Recent advances in non-Markovian processes \citep{williams2019} and temporal motifs \citep{paranjape2017} might provide useful starting points.

Our results also have practical implications. The explosive growth patterns suggest that resource allocation and infrastructure planning for collaborative systems should anticipate super-linear scaling rather than linear growth. The strong influence of external events highlights the importance of considering broader societal factors in institutional planning and policy-making.

The limitations of our study should be noted. While our datasets span long time periods, they represent only two specific types of collaboration networks. Investigation of other long-term social networks would help establish the generality of our findings. Additionally, our analysis focuses on structural properties; future work might explore the evolution of functional characteristics or information flow patterns.

These findings constitute a significant step toward understanding the long-term evolution of social networks, but they also reveal how much remains to be learned. The challenge ahead lies in developing new theoretical frameworks that can capture the rich dynamics we observe while maintaining sufficient simplicity to provide useful insights. Such frameworks will be essential for understanding and managing the increasingly connected collaborative systems that characterise modern society.

\section{Conclusions}

Through detailed analysis of two major collaboration networks—the Microsoft Academic Graph (1800-2020) and Internet Movie Database (1900-2020)—this study observes properties that may contribute to the development of existing network theory. Both networks show explosive super-linear growth that persists without saturation. The academic network exhibits two distinct growth regimes, with power-law exponents increasing from $\alpha_1 \approx 2.3$ to $\alpha_2 \approx 3.1$ after 1950. This dramatic expansion exceeds demographic scaling, with network growth outpacing population increase by several orders of magnitude.

A second fundamental discovery concerns the temporal coupling of network processes. Contrary to prevailing assumptions in network theory, node and edge dynamics maintain stable timescale ratios ($\tau_N/\tau_E$ between 2 and 3) throughout their evolution. Both timescales decrease monotonically, indicating synchronised acceleration across all network processes. This persistent coupling suggests a need to examine theoretical frameworks that assume strong separation between these timescales and suggests the need for new mathematical approaches.

The networks appear to exhibit environmental coupling, which can be observed in their node dynamics. Major historical events leave distinct signatures in network evolution, with academic collaboration showing heightened sensitivity to external influences compared to entertainment networks. This coupling manifests primarily in node addition and removal processes, while edge dynamics exhibit greater resilience, indicating distinct governing mechanisms for these processes.

Analysis of inter-collaboration waiting times reveals characteristic patterns that distinguish academic and entertainment domains. Both networks exhibit power-law distributions whose indices evolve while maintaining domain-specific ranges ($\gamma \approx 2.3$ to $1.6$ for academia, $\gamma \approx 2.6$ to $2.1$ for entertainment). The flattening of academic distributions suggests increasing diversity in research timescales, while entertainment's steeper distributions reflect more structured professional rhythms. The persistence of these functional forms amid explosive growth indicates fundamental constraints on human collaborative behaviour.

These findings suggest opportunities for refining current network evolution models. Future theoretical frameworks must accommodate sustained super-linear growth while incorporating multiple coupled timescales. They must account for environmental influences while explaining the remarkable stability of local network properties during periods of global acceleration. This presents an interesting balance between representing these dynamics while maintaining mathematical tractability.

The practical implications of these findings extend to institutional planning and policy development. organisations must prepare for super-linear rather than linear growth in both infrastructure and resource allocation, particularly in academic contexts where environmental coupling is most pronounced. Future research should investigate whether these properties represent universal features of large-scale human collaboration by extending this analysis to other types of social networks over similar temporal spans.

\begin{acks}
This research was carried out at Rinna K.K., Tokyo, Japan.
\end{acks}
\bibliographystyle{SageH}
\bibliography{paper1}

\end{document}